\documentclass[aps,prb,twocolumn]{revtex4-1}
\usepackage{latexsym}
\usepackage{float}
\usepackage{subcaption}
\usepackage{graphicx}
\usepackage{dcolumn}
\usepackage{bm}
\usepackage{amssymb}
\usepackage{amsmath}
\usepackage[space]{grffile}
\usepackage{color}

\begin{document}

\title{Transport and spectral signatures of transient fluctuating superfluids in the absence of long-range order}
\date{\today}
\begin{abstract}
Results are presented for the quench dynamics of a clean and interacting electron system, where the quench involves varying 
the strength of the attractive interaction along arbitrary quench trajectories.
The initial state before the quench is assumed to be a normal electron
gas, and
the dynamics is studied in a regime where long-range order is absent, but nonequilibrium superconducting fluctuations need to be accounted for.
A quantum kinetic equation using a two-particle irreducible formalism is derived. Conservation of
energy, particle-number, and momentum emerge naturally, with the conserved currents depending on both the electron
Green's functions and the Green's functions for the superconducting fluctuations. The quantum kinetic equation is employed to derive 
a kinetic equation for the current,
and the transient optical conductivity relevant to pump-probe spectroscopy is studied. The general structure of the kinetic equation for the
current is also justified by a phenomenological approach that assumes model-F in the
Halperin-Hohenberg classification, corresponding to a non-conserved order-parameter coupled to a conserved density.
Linear response conductivity and the diffusion coefficients in thermal equilibrium are also derived,
and connections with Aslamazov-Larkin fluctuation corrections are highlighted.
Results are also presented for the time-evolution of the local density of states. It is shown that Andreev scattering processes
result in an enhanced density of states at low frequencies. For a quench trajectory corresponding to
a sudden quench to the critical point, the density of states is shown to grow in a manner
where the time after the quench plays the role of the inverse detuning from the critical point.
\end{abstract}

\author{Yonah Lemonik}
\author{Aditi Mitra}

\affiliation{Center for Quantum Phenomena, Department of Physics, New York University, 726 Broadway, New York, New York, 10003, USA}
\maketitle


\section{Introduction}\label{sec:intro}

Recent years have seen impressive advances in ultrafast measurements of strongly correlated
systems~\cite{Smallwood12,Averitt14,Smallwood14,Wall18,Hsieh18,McIver18,Mitrano18,Murnane18}.
In these experiments,
a pump field strongly perturbs the system, while a weak probe field studies the eventual
time-evolution over time scales that can range from femto-seconds to nano-seconds. Thus one can study
the full nonequilibrium dynamics of the system from short times, to its eventual thermalization at long times. Moreover,
the access to probe fields ranging from x-rays to mid-infrared allows one to probe dynamics from
short length and time scales to longer length and time scales associated with collective modes.

Among the various pump-probe studies, notable examples are experiments that show
the appearance of highly conducting, superconducting-like
states, that can persist from a few to hundreds of pico-seconds~\cite{Fausti11,Mitrano15,Mitrano17,Averitt19}. The microscopic
origin of this physics is not fully understood. Explanations range from destabilization by the pump laser of a competing order
in favor of superconductivity~\cite{Mathey16,Cavalleri18},
to selective pumping of phonon modes that can engineer the effective electronic degrees of freedom so as to
enhance the attractive Hubbard-U, and hence \(T_c\)~\cite{Knap16,Sentef16,Sentef17b,Kennes17,Werner17}. There are also theoretical
studies that argue that the metastable highly conducting state observed in experiments may not have a superconducting origin~\cite{Aleiner18}.

The resulting superconducting-like states are transient in nature,
where traditional measurements of superconductivity, such as the Meissner effect~\cite{Tinkhambook}, cannot be performed.
On the other hand, time-resolved transport and angle-resolved photo-emission spectroscopy (tr-ARPES) are more relevant probes for such transient states.
Thus, a microscopic treatment that makes predictions for transient conductivity and spectral features of
a highly nonequilibrium system, accounting for dynamics of collective modes, is needed.
This is the goal of the paper.

In what follows, we model the
pump field as a quantum quench~\cite{Calabrese06,AMreview} in that it perturbs a microscopic parameter of the system,
such as the strength of the attractive Hubbard-U.
We assume that, as in the experiments~\cite{Mitrano15}, the electronic
system is initially in the normal phase. The quench involves tuning the magnitude of the attractive Hubbard-U along different
trajectories. How superconducting fluctuations develop in time, and how they affect
transient transport and spectral properties are studied. Denoting the superconducting order-parameter as \(\Delta\),
we assume that all throughout the dynamics, the system
never develops true long range order, \(\langle \Delta(t)\rangle=0\). Thus our approach is
complementary to other studies where the starting point is usually a fully ordered superconducting state, and
the mean-field dynamics of the order-parameter \(\langle \Delta(t)\rangle\)
is probed~\cite{Yuzbashyan06,Barankov06,Capone15,Yuzbashyan15,Foster17,Yuzbashyan19}.
In our case, the non-trivial dynamics appears in the fluctuations \(\langle \Delta(t) \Delta(t')\rangle\) which
we study using a two particle irreducible (2PI) formalism. We justify the selection of diagrams using a $1/N$ approach where
\(N\) denotes an orbital degree of freedom. The resulting choice of diagrams is equivalent to a self-consistent
random-phase-approximation (RPA) in the particle-particle channel. The physical meaning of the RPA is that the superconducting
fluctuations are weakly interacting with each other. 

While in this paper we study a clean system, a complementary study of the transient
optical conductivity of a disordered system appears in Ref.~\onlinecite{Lemonik18}, where a Kubo formalism approach was
employed, and the quench dynamics arising
from fluctuation corrections of the Aslamazov-Larkin (AL)~\cite{AL68-1,AL68-2} and Maki-Thompson (MT)~\cite{Maki68-1,Maki68-2,Thompson68}
type was explored.
We avoid a Kubo-formalism approach in this paper because conserving approximations are harder to make, as compared to
a quantum kinetic equation approach. This is an observation well known from other studies for transport in thermal
equilibrium performed for example in the
particle-hole channel~\cite{Catelani05}. Since we are in a disorder-free system, our quantum kinetic equations
conserve momentum. Thus, to obtain any non-trivial current dynamics, we have to break Galilean invariance.
We do this through the underlying lattice dispersion, while neglecting Umpklapp processes.

Complementary studies of transient conductivity involve phenomenological time-dependent Ginzburg-Landau (TDGL)
theories~\cite{Kennes17a,Hoshino19}, and microscopic approaches based on a Bardeen-Cooper-Schrieffer (BCS) mean-field treatment of the
order-parameter dynamics~\cite{Kennesprb17}. A microscopic mean field approach has also
been used to study transient spectral densities~\cite{Moore19}, while
nonequilibrium dynamical mean field theory has been used to study transient spectral properties of a BCS superconductor~\cite{Stahl18}.

The outline of the paper is as follows. In Section~\ref{sec:Model} we present the model. In this section we also outline a
qualitative analysis for the transient conductivity employing a classical Model-F~\cite{HH77} scenario for
the fluctuation dynamics. In Section~\ref{sec:EOM}
we derive the quantum kinetic equations using a two particle irreducible (2PI) formalism, and outline how the transient
dynamics respects all the conservation laws (particle, energy, momentum) of the system. In Section~\ref{sec:classical_reduction} we simplify
the 2PI quantum kinetic equation to that of an effective classical equation for
the decay of the current employing a separation of time-scales. In the process we justify
the model-F scenario discussed in Section~\ref{sec:Model}. In Section~\ref{sec:eqtr} we assume thermal equilibrium, and derive
the conductivity and diffusion constants. We discuss these in the context of Aslamazov-Larkin (AL) and Maki-Thompson (MT) corrections~\cite{Larkin00}.
In Section~\ref{sec:solving} we present results for the transient conductivity for
some representative quench profiles.
In section~\ref{sec:spectral} we present results for the transient local density of states, and finally in Section~\ref{sec:Concl}
we present our conclusions. Many intermediate steps of the derivations are relegated to the appendices.

\section{Model} \label{sec:Model}

The Hamiltonian describes fermions with spin and an additional orbital degree on a regular lattice in $d$ spatial dimensions.
Fermions are created by the operator $c^\dagger_{\sigma n}(r_i)$, where $\sigma = \pm 1$ labels spin, $n = 1\ldots N$ labels the
orbital index and $r_i$
gives the lattice site. These have the Fourier transform $\tilde{c}_{\sigma n} (k) = \sum_{r_i} e^{i \vec{k} \cdot \vec{r}_i}c_{\sigma n} (r_i)$.
The dynamics of the fermions is governed by the Hamiltonian:
\begin{align}
H(t) &\equiv H_0 +  H_{\rm int}(t),\\
H_0 &\equiv \int_{\rm BZ} \frac{d^d k}{(2\pi)^d} \sum_{\sigma,n} (\epsilon_k -\mu) \tilde{c}^\dagger_{\sigma n}(k) \tilde{c}_{\sigma n}(k),\\
H_{\rm int} &\equiv \frac{U(t)}{N} \sum_{r_i,n,m,\sigma}c^\dagger_{\sigma,n}(r_i)c^\dagger_{-\sigma,n}(r_i)c_{-\sigma,m}(r_i)c_{\sigma,m}(r_i)\label{H},
\end{align}
where $\epsilon_k$ is the dispersion, and $\mu$ is the chemical potential.
We will assume for simplicity that at this chemical potential there is only a single Fermi surface.
We explicitly allow for the interaction $U(t)$ to vary with time.

In addition, to probe current dynamics, we introduce an electric field \(\vec{E}=-\partial_t\vec{A}\) by minimal
coupling \(\vec{k} \rightarrow \vec{k} -q \vec{A}\). Accounting for the charge of the electron \(q=-e\),
this appears in the Hamiltonian via \(\vec{k} \rightarrow \vec{k} + e\vec{A}\), or equivalently,
\(\vec{\nabla}/i \rightarrow \vec{\nabla}/i + e \vec{A}\).
In the microscopic derivation we will set \(\hbar =1\).

In thermal equilibrium at a temperature $T$, and for spatial dimension $d>2$, the above system becomes superconducting below
a critical temperature $T< T_C$. At $d=2$,
there is a Berezenskii-Kosterlitz-Thouless (BKT) transition at $T<T_{\rm BKT}$ where the system shows only quasi-long range order.
For $T>T_{\rm BKT}$, although the system has no long range order, yet superconducting fluctuations play a key role in transport.
The lower the spatial dimensions, the more important the role of fluctuations. In particular for spatial dimension $d=2$, well
known results for the optical conductivity exist for a strongly disordered system~\cite{Larkin00}. While our derivation is
valid in any spatial dimension, we will present results for spatial dimensions \(d=2\) where fluctuation effects are most pronounced. The experiments
also study either a layered superconducting system, or three dimensional systems where the pump field is effectively
a surface perturbation. This  makes the spatial dimension \(d=2\) also experimentally relevant.

Before we go into details, we note that our microscopic treatment leads to  the following expression (reported in Ref.~\onlinecite{Lemonik17b})
for the dynamics of the current \(j^i\), generated by an
applied electric field \(E^i\),
\begin{align}
&\partial_t j^i(t) - E^i(t) = -\frac{1}{\tau_{r}}\biggl[A(t) j^i(t)\nonumber\\
&- \alpha \int_0^t dt' \biggl\{B (t,t') j^i(t') + C(t,t')E^i(t')\biggr\}\biggr].\label{jke1}
\end{align}
Above \(\tau_r,A,B,C\) are kinetic coefficients. We have set \(e=1,\rho/\bar{m}=1\), where \(\rho\) is the
electron density, and \(\bar{m}\) is the electron effective mass. When \(B=C=0\), the above equation simply implies
a time-dependent Drude scattering rate \(\tau_r^{-1}A(t)\), which in our model arises because electrons scatter off
superconducting fluctuations whose density is changing with time due to the quench. The appearance of memory terms
\(B,C\) is non-trivial, and before deriving them, we will justify the
appearance of these memory terms through a phenomenological approach in the next subsection.

\subsection{Model F dynamics}
\label{sec:Model_F}

In a finite temperature phase transition the dynamics of the fluctuations may be understood by treating them as classical stochastic fields.
This will be seen by direct calculation later in the paper. However for now,
we seek to write a Langevin equation for the (classical) field $\Delta$.
In the theory of dynamical critical phenomena, we must account not only for the fluctuating mode, $\Delta$, but also for any conserved quantities
which are coupled to $\Delta$. The pairing field $\Delta$ couples directly to the density \(\rho\), as given by the commutator
$\left[ \Delta, \rho \right]  = 2\Delta$.
This can be translated into classical terms by the usual prescription $\left[\cdot,\cdot\right]/i\hbar\rightarrow \{ \cdot, \cdot\}$ where
$\{ \cdot, \cdot\}$ is the classical Poisson bracket.

In the framework of dynamical critical phenomena we specify the static free energy (see Ref.~\onlinecite{HH77}, Eq.~(6.3))
\begin{align}
&\mathcal{F}\left[\Delta,\rho\right] \equiv \int d^d r \left[  e\Phi\delta\rho +
\frac{1}{2 C}(\delta\rho)^2 + r(t)|\Delta|^2 \right. \nonumber\\
&\left. + \frac{1}{2M}|(i\vec{\nabla} - 2e\vec{A}) \Delta|^2 + u |\Delta|^4 + \gamma_0 (\delta\rho)|\Delta|^2\right].\label{FmodelF}
\end{align}
Here $\delta\rho$ is density minus the equilibrium density, $\Phi$ is the scalar potential,
$C$ is the compressibility, $r(t)$ is the time dependent
detuning from the critical point, $\vec{A}(t)$ is an external vector potential, and $u$, $\gamma_0$ are phenomenological parameters.
Note that model E~\cite{HH77} has $\gamma_0=0$, while model F has $\gamma_0\neq 0$.

The field $\Delta$ then has the equation of motion,
\begin{align}
\left(\frac{\partial \Delta}{\partial t} - \left\{\Delta, \rho \right\}\frac{\partial \mathcal{F}}{\partial \rho} \right)
&=
-\Gamma_\Delta
	\frac{\partial \mathcal{F}}{\partial \Delta^*}
	+ \xi_\Delta.
\end{align}
Using Eq.~\eqref{FmodelF}, we obtain,
\begin{align}
&\left(
	\frac{\partial\Delta}{\partial t}
	+ 2i\Delta\left[
		e\Phi
		+ \frac{\delta\rho}{C}
		+ \gamma_0|\Delta|^2
	\right]
\right)
=
-\Gamma_\Delta \biggl[r(t) + \gamma_0\delta\rho \nonumber\\
& + 2 u |\Delta|^2 + \frac{1}{2M}(i\nabla^i -2eA^i)^2\biggr]\Delta +\xi_\Delta,\label{deldyn}
\end{align}
where ${\xi}_\Delta$ is a Gaussian white noise obeying $\langle\xi_\Delta(r,t) \xi_{\Delta^*}(r',t')\rangle = 2T \Gamma_\Delta \delta(r-r')\delta(t-t')$, and
$\Gamma_{\Delta}$ is a dimensionless phenomenological damping.

We will denote the total current by a dissipative component $j_{\rm dis}$, and a superfluid component $j_{\rm sc}$, $j^i = j^i_{\rm sc} + j^i_{\rm dis}$.
The equation of motion for the density $\rho$ is given by,
\begin{align}
\frac{\partial \delta\rho
	}{
\partial t}&
	- \left\{\rho,\Delta\right\}\frac{\partial \mathcal{F}}{\partial\Delta}
	- \left\{\rho,\Delta^*\right\}\frac{\partial \mathcal{F}}{\partial\Delta^*}
	=
	\vec{\nabla}\cdot\vec{j}_{\rm dis}\label{F1},
\end{align}
where the right hand side (rhs) can be interpreted as a definition of \(j_{\rm dis}\).
Under assumptions that the current may be written as a gradient of a scalar, and that it reaches a well defined steady state
in the presence of a dc electric field,
\begin{align}
\vec{j}_{\rm dis} &=
	-\Gamma_{j_{\rm dis}}\left( \vec{\nabla}
	\frac{\partial\mathcal{F}}
	{\partial \rho}
	+ e\frac{\partial \vec{A}}{\partial t}\right)
	+  \vec{\xi}_{j_{\rm dis}}.\label{F3}
\end{align}
Above, $\vec{\xi}_{j_{\rm dis}}$ is a Gaussian white noise, whose strength is now controlled by $T\Gamma_{j_{\rm dis}}$, where
$\Gamma_{j_{\rm dis}}$ is another phenomenological dissipation. Using the fact that $\partial_t \vec{A}=-\vec{E}$, the steady-state dissipative current
for a spatially homogeneous system is (setting \(e=1\))
\begin{equation}
j^i_{\rm dis} = \Gamma_{j_{\rm dis}} E^i.\label{jdis}
\end{equation}
Now using,
\begin{align}
\left\{\rho,\Delta\right\}\frac{\partial \mathcal{F}}{\partial\Delta}
	&+ \left\{\rho,\Delta^*\right\}\frac{\partial \mathcal{F}}{\partial\Delta^*}
=
 	\frac{i}{M}\biggl[
		\Delta (i\nabla^i + 2 e A^i)^2 \Delta^*\nonumber\\
&		- \Delta^* (i\nabla^i - 2 e A^i)^2 \Delta
	\biggr]
= \vec{\nabla}\cdot \vec{j}_{\rm sc}\label{F2},
\end{align}
we find that the definition of the superfluid current naturally emerges,
\begin{align}
\vec{j}_{\rm sc} &= \frac{2}{M}\text{Im}\left[
	\Delta (\vec{\nabla} - 2i e \vec{A})\Delta^*\right].
\end{align}
Substituting Eq.~\eqref{F2} in Eq.~\eqref{F1}, we have,
\begin{align}
&\frac{\partial \delta\rho}{\partial t} = \vec{\nabla}\cdot \vec{j}_{\rm sc} + \vec{\nabla}\cdot \vec{j}_{\rm dis}.
\end{align}

We now relate these equations to our equations of motion Eq.~\eqref{jke1}.
The expectation value for the supercurrent is
\begin{align}
\vec{j}_{\rm sc} &= \frac{2}{M}\text{Im}\langle \Delta(x) \vec{\nabla} \Delta^*(x)\rangle
\nonumber \\
	&= \frac{2}{M} \int \frac{d^d q}{(2\pi)^d} \vec{q} \langle {\Delta}_q{\Delta}_q^*\rangle
\nonumber \\
	& = 2\int\!\! \frac{d^dq}{(2\pi)^d}\, \frac{\vec{q}}{M}  F(q,t),
\end{align}
where $F(q)$ is the equal time $\Delta_q \Delta_q^*$ correlator at momentum $q$. Now we show that the integral or memory term
proportional to $\alpha$ in Eq.~\eqref{jke1} comes from a term of the form
\begin{equation}
	\propto \int\!\! d^dq\, \frac{\vec{q}}{M}  \partial_t F(q,t),
\end{equation}
and that this term is precisely $\partial_t j_{\rm sc}$. The remainder can therefore be identified with the dissipative current giving an equation
\begin{equation}
\partial_t j^i_{\rm dis} - E^i(t) = -\tau_r^{-1}A(t) j^i_{\rm dis}.
\end{equation}
This should be compared to the expectation value for the dissipative current Eq.~\eqref{jdis}.
(Note that the damping coefficient \(A\) is not to be confused with the vector potential \(A^i\)).
Thus in the dc limit of the kinetic equation, we identify $\Gamma_{j_{\rm dis}} = \tau_r A^{-1}$.
Note that the model F dynamics are less general in the sense
that they assume this dc limit, and therefore that all perturbations are slow compared with
$\Gamma_{j_{\rm dis}}$. The kinetic equation does not make this assumption.

Now let us evaluate $\partial_t j_{\rm sc}$. We simplify the generalized time-dependent Ginzburg-Landau theory
in Eq.~\eqref{deldyn} by going into Fourier space, and dropping all non-linear terms,
\begin{align}
\frac{\partial\Delta_q}{\partial t}
&=
-\Gamma_\Delta \left[r(t) + \frac{1}{2M}(q^i +2eA^i(t))^2+\gamma_0 \delta \rho_q(t)\right]\Delta_q \nonumber\\
&+\xi_\Delta.
\end{align}
Above we have encoded the effect of the electric field in two places, one in the minimal coupling $\vec{q} \rightarrow \vec{q}+2e\vec{A}$,
and second in the change in the electron density at momentum $q$, where by symmetry,
\begin{align}
\delta \rho_q(t) = c j^i(t) q^i,
\end{align}
where \(c\) is a phenomenological constant. Going forward, we absorb this constant into a redefinition of \(\gamma_0\) appearing
in the combination \(\gamma_0\delta\rho_q(t)\).

Defining,
\begin{eqnarray}
\lambda_q(t) = \Gamma_{\Delta}\left[r(t) + \frac{q^2}{2M}\right]=\Gamma_{\Delta}\epsilon_q(t),
\end{eqnarray}
in the absence of an electric field, the solution is
\begin{align}
\Delta_q(t) = \int_0^t dt' \xi_{\Delta}(t') e^{-\int_{t'}^tdt''\lambda_q(t'')}.
\end{align}
Above we have adopted boundary conditions where the superconducting order-parameter and fluctuations are zero at \(t=0\). This is because we are
interested in quenches where there are initially \(t\leq 0\) no attractive interactions, with these being
switched on at some arbitrary rate from \(t>0\).

The fluctuations involve averaging over noise, giving,
\begin{align}
F(q,E=0,t)=\langle \Delta_q\Delta_q^*\rangle = 2\Gamma_{\Delta} T\int_0^t dt' e^{-2\int_{t'}^tdt''\lambda_q(t'')}.\label{Fqsola}
\end{align}
In the presence of an electric field, and to leading order in it, $\lambda_q$ changes to $\lambda_q +\delta\lambda_q$, with
\begin{align}
\delta \lambda_q(s) = \frac{\Gamma_{\Delta}}{M}(2e)\vec{q}\cdot\int_{s}^tdt'\vec{E}(t') +\Gamma_{\Delta}\gamma_0q^i j^i(s),\label{delamdef}
\end{align}
where the first term is due to the order-parameter being charged, and the second term is due to the coupling of the order-parameter
to the normal electrons whose density is perturbed by the electric field. We have also used that the electric field is
$\vec{E} =-\partial_t \vec{A}$, so that \(\vec{A}(t)= -\int _s^t dt' \vec{E}(t')\).

Expanding Eq.~\eqref{Fqsola} in $\delta \lambda_q$, which is equivalent to expanding in the electric field,
\begin{align}
&F(q,t) \simeq  2\Gamma_{\Delta} T\int_0^t dt' e^{-2\int_{t'}^tdt''\lambda_q(t'')}\biggl[1
-2\int_{t'}^tdu \delta\lambda_q(u)\biggr].\nonumber
\end{align}
Splitting the integral in the exponent, $\int_{t'}^t dt''= \int_{u}^tdt'' + \int_{t'}^u dt''$, and using Eq.~\eqref{Fqsola},
we arrive at the following expression for the superconducting fluctuations at momentum \(q\), to leading non-zero order in the
applied electric field, for an arbitrary time-dependent detuning \(r(t)\),
\begin{align}
&F(q,t) = F(q,E=0,t) +\delta F(q,t),\nonumber\\
&\delta F(q,t) = -2\int_0^t du \delta \lambda_q(u)e^{-2\int_{u}^tdt''\lambda_q(t'')}F(q,E=0,u)\label{Fqe}.
\end{align}

\subsubsection{Aslamazov-Larkin (AL)}
We now briefly show how one may recover the familiar AL conductivity in thermal equilibrium. For this it suffices to
revert to model E by setting  \(\gamma_0=0\). For a static electric field, and a system in thermal
equilibrium, all couplings are time-independent. Thus Eq.~\eqref{Fqe} becomes,
\begin{align}
\delta F(q,t)= -2 F(q) \int_0^{t}du \delta\lambda_q(u)e^{-2(t-u)\lambda_q}.
\end{align}
Using Eq.~\eqref{delamdef} for \(\gamma_0=0\), \(\delta \lambda_q(u) = \frac{\Gamma_{\Delta}}{M} (2 e)(t-u) \vec{q}\cdot \vec{E}\).
Taking the long time \(t\rightarrow \infty\) limit, and employing the equilibrium expression for \(F(q)=T/\epsilon_q\), where \(\epsilon_q= r+q^2/2M\),
the change in the density of superconducting fluctuations due to the applied electric field is,
\begin{align}
\delta F_q = \Gamma_{\Delta}^{-1}\frac{T}{\epsilon_q^3}\frac{\vec{q}}{M}\cdot e\vec{E}.
\end{align}
The current is
\begin{align}
j^i = 2 e \sum_q \frac{q^i}{M}\delta F(q) = \Gamma_{\Delta}^{-1}2e^2\sum_q\frac{T}{\epsilon_q^3}\frac{q^iq^j}{M^2} E^j,
\end{align}
leading to the AL conductivity in \(d\) dimensions,
\begin{align}
\sigma_{\rm AL} = 2e^2\Gamma_{\Delta}^{-1}T\int \frac{d^dq}{(2\pi)^d}\frac{q^2}{d M^2}\frac{1}{\left(r + q^2/2M\right)^3}.
\end{align}
In \(d=2\), this reduces to
\begin{align}
\sigma_{\rm AL}^{d=2} = e^2\frac{1}{2\pi\Gamma_{\Delta}}\frac{T}{r}.
\end{align}
A microscopic treatment involving electrons in a disordered potential shows that~\cite{Larkin00} \(\Gamma_{\Delta}^{-1}=\pi/8\),
giving the well known expression,
\(\sigma_{\rm AL}^{d=2} = \frac{e^2}{16}\frac{T}{r}\). In our disorder-free model, \(\Gamma_{\Delta}\) takes a different value. 

\subsubsection{Kinetic equation for the current}
We now return to the derivation of the dynamics of the current in model F.
For the current dynamics we need to differentiate Eq.~\eqref{Fqe} with time,
\begin{align}
&\partial_t F(q,E,t) =  \partial_t F(q,E=0,t)- 2\delta \lambda_q(t) F(q,E=0,t) \nonumber\\
&+4\lambda_q(t)\int_0^t du
\delta \lambda_q(u)e^{-2\int_{u}^tdt''\lambda_q(t'')}F(q,E=0,u).
\end{align}
The first term on the rhs, being symmetric in momentum space, does not
contribute to the current. The second term on the rhs simply provides a time-dependent Drude scattering rate. It is 
the last term, namely the memory term proportional to the electric field, that is unique to having
long lived superconducting fluctuations. We focus only on this term, and using Eq.~\eqref{delamdef}, write it as,
\begin{align}
&\partial_t j_{\rm sc}^i(t) = 2\int\!\! \frac{d^dq}{(2\pi)^d}\, \frac{q^i}{M}  \partial_t F(q,t)\nonumber\\
&=\frac{8\Gamma_{\Delta}}{M}\int \frac{d^dq}{(2\pi)^d} q^i q^j \lambda_q(t)\int_0^t du \biggl[\gamma_0 j^j(u) \nonumber\\
&+\frac{2 e}{M} \int_{u}^tdt'{E}^j(t')\biggr]
e^{-2\int_{u}^tdt''\lambda_q(t'')}F(q,u).\label{jdyn2}
\end{align}
Above, we have dropped the \(E=0\) label in \(F\).
In the second term, we will find it convenient to replace the time integrals as follows
\(\int_0^tdu\int_u^tdt' \rightarrow \int_0^tdt'\int_0^{t'}du\).

Comparing equations \eqref{jke1} and \eqref{jdyn2}, we identify,
\begin{align}
\tau_r^{-1}\alpha B(t,t') &= \frac{8\Gamma_{\Delta}\gamma_0}{d M}\int \frac{d^dq}{(2\pi)^d} q^2 \lambda_q(t)\nonumber\\
&\times e^{-2\int_{t'}^tdt''\lambda_q(t'')}F(q,t')\label{BmodelF}.
\end{align}
The second memory term may
be identified as,
\begin{align}
\tau_r^{-1}\alpha C(t,t') &=\frac{8\Gamma_{\Delta}}{d M^2}(2e)\int \frac{d^dq}{(2\pi)^d} q^2\lambda_q(t)\nonumber\\
&\times \int_0^{t'} du e^{-2\int_{u}^tdt''\lambda_q(t'')}F(q,u).\label{CmodelF}
\end{align}
Comparing Eq.~\eqref{BmodelF} and Eq.~\eqref{CmodelF}, we obtain,
\begin{align}
C(t,t') = \frac{2e}{M\gamma_0}\int_0^{t'} ds B(t,s).
\end{align}
The microscopic treatment that follows not only justifies model F, it also provides an explicit calculation for
the parameters $\tau_r^{-1},A(t),\alpha,\Gamma_{\Delta}, \Gamma_{j_{\rm dis}},M, r,\gamma_0$.

\section{$2$PI Equations of Motion}
\label{sec:EOM}

\subsection{Properties of $2$PI formalism}
\label{sec:2PI_properties}
We briefly recapitulate the $2$PI formalism here. A detailed explanation can be found for example in
Refs~\onlinecite{CornwallJackiwTomboulis,Ivanov98,AST61,BergesRev}.
The $2$PI  formalism begins with the Keldysh action $S$ for the Hamiltonian $H$. This is written in terms of Grassmann fields
$\psi_{\sigma,n,\pm}(r,t)$,$\bar{\psi}_{\sigma,n,\pm}(r,t)$  where $\sigma$ labels the spin, $n$ labels the orbital quantum number,
and $\pm$ label the two branches of the Keldysh contour~\cite{Kamenevbook2011}. In terms of these, the Keldysh action is given as
\begin{align}
&S_K[\psi,\bar{\psi}] \equiv \int\!  dt\, \bigg\{\,\,
i\!\!\!\!\sum_{\sigma =\uparrow,\downarrow; n,r}\!\!\!\!  \bar{\psi}_{\sigma n +}(r,t) \frac{\partial}{\partial t}\psi_{\sigma n +}(r,t)\nonumber\\
&- H\left[\bar{\psi}_+(t),\psi_+(t)\right] \,\,
-i\!\!\!\!\sum_{\sigma =\uparrow,\downarrow; n,r}\!\!\!\!  \bar{\psi}_{\sigma n -}(r,t) \frac{\partial}{\partial t}\psi_{\sigma n -}(r,t)\nonumber\\
&+ H\left[\bar{\psi}_-(t),\psi_-(t)\right]\bigg\},
\end{align}
where $H\left[\bar{\psi}_\pm(t),\psi_\pm(t)\right]$ indicates the substitution of $\psi_\pm(t)$ for $c(t)$ in the Hamiltonian,
Eq.~\eqref{H}, in the obvious way. Now we consider a classical field $h$ that couples to a general bilinear of the Grassmann field,
and is therefore given by $h_{\sigma\,n\,p}^{\sigma'n'p'}(r,t,r',t')$, where $p$,$p'$ run over $\pm$.
In order to simplify notation, we combine the five indices, spin, orbital, Keldysh, position and time into a single vector index,
so that the Grassman field becomes a vector $\psi_i$ and the source field a matrix $h_{ij}$.

From the action and the source field, we form a generating functional $\mathcal{I}[h]$:
\begin{equation}
\mathcal{I}[h] = \int \mathcal{D}\left[\bar{\psi},\psi\right]\exp\bigg[i S_K[\bar{\psi},\psi]
+ i\sum_{ij} \bar{\psi}_i h_{ij}\psi_j\bigg].
\end{equation}
The first derivative of $\mathcal{I}$ is precisely the Green's function in the presence of the source field $h$
\begin{equation}
G_{ij}[h] = \frac{\delta \mathcal{I}}{\delta h_{ji}} = i\int\mathcal{D}\left[\bar{\psi},\psi\right]\,\,\left(\bar{\psi}_j \psi_i\right) e^{iS_K+i\bar{\psi}\cdot h\cdot\psi}.
\label{eq:G_in_terms_of_h}
\end{equation}

In order to work with the physical Green's function, rather than the unphysical source field $h$, we perform a Lagrange inversion.
First we invert Eq.~\eqref{eq:G_in_terms_of_h} which gives $G$ as a function of $h$, to implicitly
define $h$ as a function $G$. Then we construct the Lagrange transformed functional $\mathcal{W}$,
\begin{equation}
\mathcal{W}[G] = \mathcal{I}[h(G)] - h(G) G.
\end{equation}
This has the property that,
\begin{equation}
\frac{ \delta \mathcal{W}}{\delta G} = -h(G).
\end{equation}
In particular in the physical situation when $h = 0$ we have that $\delta \mathcal{W} / \delta G = 0$.
Thus given the functional $\mathcal{W}$, we have reduced the problem of calculating $G$ to a minimization problem.

The functional $\mathcal{W}$ can be calculated in terms of a perturbation series in $U$, giving
\begin{equation}
\mathcal{W}[G] = \frac{i}{2}\text{Tr}\ln(G^{-1}) + \frac{i}{2}\text{Tr}\left[g^{-1}G\right]   + \Gamma_2[G],
\end{equation}
here the Tr operates on the entire combined index space, and  $g^{-1}$ is the bare electron Green's function.
The functional $\Gamma_2[G]$ is constructed as follows. First the Feynman
rules for the Hamiltonian $H$ are written down. In the present case, this is a single fermionic line and a four-fermion
interaction.
Then all two-particle irreducible bubble diagrams are drawn. A diagram is two particle irreducible (2PI)
if it cannot be disconnected by deleting up
to two fermionic lines. The functional $\Gamma_2[G]$ is given by the sum of all the diagrams.
Lastly the diagrams are interpreted as in
usual diagrammatic perturbation theory except that the fermionic line is not the bare Green's function $g$,
but instead the full Green's function $G$.

The end result is the minimization condition
\begin{align}
\frac{\delta \mathcal{W}}{\delta G} &= -\frac{i}{2}G^{-1} + \frac{i}{2}g^{-1} + \frac{\delta \Gamma_2}{\delta G} = 0,\\
\rightarrow G^{-1} &= g^{-1} -\Sigma,\, \Sigma\left[G\right] = 2i\delta \Gamma_2 /\delta G.
\label{eq:full_dyson_eq}
\end{align}
This is the Dyson equation for the Green's function $G$ where the self energy $\Sigma[G] \equiv 2i\delta \Gamma_2 /\delta G$ is a
self-consistent function of $G$. As no approximations have been made  Eq.~\eqref{eq:full_dyson_eq} is a restatement of
the original many-body problem, and
clearly cannot be solved.  The advantage of the formalism is that we may replace  $\Gamma_{2}$ with an approximate functional $\Gamma_2'$
and still be guaranteed to preserve conservation laws
and causality, which is not the case if one directly approximates $\Sigma$.

We assume that all symmetries are maintained by the solution to Eq.~\eqref{eq:full_dyson_eq}.
Therefore the Green's function can be written in terms of the original spin, orbital, Keldysh, space, and time indices as,
\begin{align}
G_{\sigma_1,n_1,p_1;\sigma_2,n_2,p_2}\left(r_1,t_1;r_2,t_2\right) &= \delta^{\sigma_1\sigma_2}\delta^{n_1,n_2} \nonumber\\
&\times G_{p_1p_2}\left(r_1 - r_2; t_1,t_2\right),
\end{align}
so that only the Keldysh indices $p_1$, $p_2$ will be explicitly noted.
\begin{figure}[ht]
  \includegraphics[width = 0.45\textwidth]{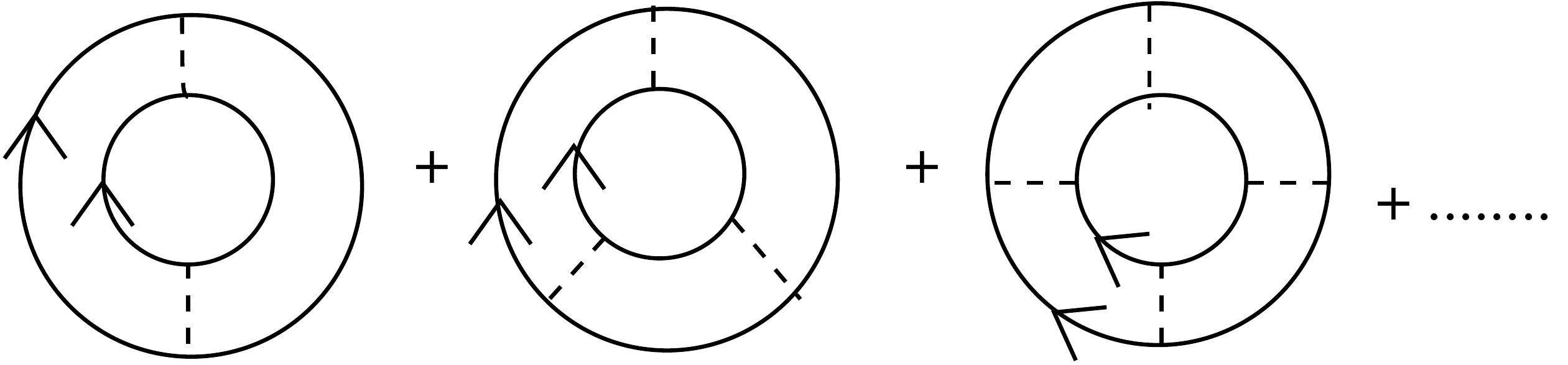}
\caption{2PI Ladder diagrams for \(\Gamma_2'\). Solid lines are electron Green's functions, while the
dashed lines are the interaction U.}\label{fig1}
\end{figure}

\subsection{RPA approximation to the 2PI generating functional}
\label{sec:RPA_2PI}

The approximation we make is the random phase approximation (RPA) in the particle-particle channel, or,
equivalently the $N\rightarrow \infty$ where $N$ is the number of orbital degrees of freedom.
That is we approximate $\Gamma_2$ by $\Gamma'_2$ where the latter include the set of all closed ladder diagrams,
\begin{gather}
\Gamma'_2 \equiv -\frac{i}{2}\Pi\cdot D \qquad D \equiv \left(U^{-1} - \Pi\right)^{-1},\\
\Pi_{p_1 p_2}\left(r_1,r_2;t_1,t_2\right) \equiv \frac{i}{2} \left[G_{p_1 p_2}\left(r_1-r_2;t_1,t_2\right)\right]^2,
\end{gather}
shown diagrammatically in Fig.~\ref{fig1}, which gives the equation,
\begin{align}
\Sigma_{p_1p_2}(r_1,t_1;r_2,t_2) &= iD_{p_1p_2}(r_1,t_1;r_2,t_2)\nonumber\\
                               &\times G_{p_2p_1}(r_2,t_2;r_1,t_1).
\end{align}
Above we have used that \(\Sigma(1,2)=2i\delta\Gamma_2/\delta G(2,1)\).
The equations may be equivalently derived from the functional for two Green's functions $G$ and $D$ given by
\begin{align}
&\mathcal{W}'[G,D]= \frac{i}{2}\text{Tr}\left[\ln(G^{-1}) + g^{-1}G +\ln(D^{-1}) + U^{-1}D\right] \nonumber\\
&\,\,\,\,\,\,\,\,\,\,\,\,\,\,\,\,\,\,\,\,\,\,\,\,\,\,\,\,\,\,+ \Gamma'_2[G,D],\\
&\Gamma'_2[G,D]\equiv -\frac{i}{2}\Pi\cdot D,\\
&\frac{\delta \mathcal{W}'}{\delta D} = -\frac{i}{2}D^{-1} + \frac{i}{2}U^{-1} - \frac{i}{2}\Pi = 0,\\
&\frac{\delta \mathcal{W}'}{\delta G} = -\frac{i}{2}G^{-1} +\frac{i}{2}g^{-1} - \frac{i}{2}\Sigma = 0.
\end{align}
This may be interpreted as the functional for electrons interacting with a fluctuating pairing field whose Green's function is $D$.
The $\Gamma'_2$ is the minimal diagram which has interaction between the fermions and the fluctuating pairing field.

\subsection{RPA equations of motion}
\label{sec:RPA_EOM}
We now proceed to analyze the RPA equations of motion. As is customary, we perform a unitary rotation of the Keldysh space,
defining $G_{K,A,R}$ by~\cite{Kamenevbook2011}
\begin{align}
G_K &\equiv \left(G_{++} + G_{+-} + G_{-+} + G_{--}\right)/2,\\
G_R &\equiv \left(G_{++} - G_{+-} + G_{-+} - G_{--}\right)/2,\\
G_A &\equiv \left(G_{++} + G_{+-} - G_{-+} - G_{--}\right)/2,\\
0 &= G_{++} - G_{+-} - G_{-+} + G_{--},
\end{align}
the last equality holding by causality. We likewise define $\Sigma_{K,A,R}, D_{K,A,R}, \Pi_{K,A,R}$. In this basis we may write the
equations of motion as (setting \(e=1\))
\begin{subequations}\label{eq:RPA_EoM}
\begin{align}
&\left[i\partial_{t_1} - \epsilon\biggl(i\vec{\nabla}_1 - \vec{A}(r_1,t_1)\biggr)\right] g_R(r_1,t_1;r_2,t_2) \nonumber\\
&= \delta(t_1-t_2),\\
&G_R^{-1} = g_R^{-1} - \Sigma_R,\\
&G_K = G_R\cdot \Sigma_K \cdot G_A,\\
&D_R^{-1} = U^{-1} -  \Pi_R, \label{Drdef1}\\
&D_K = D_R \cdot \Pi_K \cdot D_A\label{Dkdef1},
\end{align}
and where $\epsilon$ denotes the single particle dispersion, the symbol \(\cdot\) convolution, and $\Sigma, \Pi$ are given by
\begin{align}
\Sigma_R(x,y) &=  \frac{i}{2}\biggl[D_R(x,y)G_K(y,x) +  D_K(x,y)G_A(y,x)\biggr], \label{Sr} \\
\Sigma_K(x,y) &=  \frac{i}{2}\biggl[D_K(x,y)G_K(y,x) +  D_R(x,y)G_A(y,x) \nonumber\\
&+ D_A(x,y)G_R(y,x)\biggr], \\
\Pi_R(x,y) &=  \frac{i}{2}G_K(x,y)G_R(x,y),\\
\Pi_K(x,y) &=  \frac{i}{4}\left[G_K(x,y)^2+ G_R(x,y)^2+ G_A(x,y)^2\right],
\end{align}\label{SigDef}
\end{subequations}
and where $x$ and $y$ stand for combined space, and time indices. The function $G_A$ is given by $G_A = G_R^\dagger$,
where $O(x,y)^\dagger = O(y,x)^*$, and likewise for $D_A,\Sigma_A,\Pi_A$.
We note that we also have the relationship $G_K^\dagger = - G_K$

We now convert the definition of $G_K$ into a more useful form by evaluating
$G_R^{-1}\cdot G_K - G_K \cdot G_A^{-1}$. On the one hand, substituting in the definition of $G_{R,A}^{-1}$ this is,
\begin{equation}
G_R^{-1}\cdot G_K - G_K \cdot G_A^{-1}\!\!\!  = g_R^{-1}\cdot G_K - G_K \cdot g_A^{-1} - \Sigma_R\cdot G_K + G_K \cdot \Sigma_A,
\end{equation}
on the other hand substituting in $G_K = G_R\cdot \Sigma_K\cdot G_A$ we obtain
\begin{equation}
G_R^{-1}\cdot G_K - G_K \cdot G_A^{-1}  =  \Sigma_K\cdot G_A - G_R \cdot \Sigma_K.
\end{equation}
Therefore we obtain the fundamental equation
\begin{equation}
g_R^{-1}\cdot G_K - G_K \cdot g_A^{-1} = \Sigma_K\cdot G_A - G_R\cdot \Sigma_K  +\Sigma_R \cdot G_K - G_K \cdot \Sigma_A.
\end{equation}
Setting the two times equal and using the form of $g_R$ we obtain,
\begin{align}
&\partial_t iG_K(r_1,t;r_2,t) -\biggl[\epsilon\left(i\vec{\nabla}_{r_1} - \vec{A}(r_1,t)\right)  \nonumber\\
&- \epsilon\left(-i\vec{\nabla}_{r_2} - \vec{A}(r_2,t)\right)\biggr] G_K(r_1,t;r_2,t) = S(r_1t,r_2t),\nonumber\\
&S(r_1t,r_2t') \equiv \!\int\!\! d y\,\,\big\{ \Sigma_K(r_1,t;y)G_A(y,r_2,t') \nonumber\\
&+ \Sigma_R(r_1,t;y)G_K(y;r_2,t')- G_R(r_1,t;y)\Sigma_K(y;r_2,t') \nonumber\\
&- G_K(r_1,t;y)\Sigma_A(y;r_2,t') \big\}.
\label{eq:quantum_kinetic}
\end{align}
Above $y$ includes both space and time coordinates, and we have given the collision term \(S\) on the
rhs for unequal times as we will need
it to prove different conservation laws in the next section.

\subsection{Conservation laws}
\label{sec:conservation_laws}
We emphasize that the true Green's function are found by minimizing with respect to the full 2PI potential $\Gamma_2$,
whereas the functions $G$,$D$ defined above are only approximations, found by minimizing with respect to the approximate
potential $\Gamma'_2$. Therefore it is not apriori clear that there are any conserved quantities that correspond to the conserved quantities
of the true Hamiltonian. However, the existence of such quantities is guaranteed by the 2PI formalism.
Here we show that the naive expression for the conserved density
\begin{equation}
n(r,t) \equiv iG_K(r,t;r,t),\label{ndef}
\end{equation}
is correct. Setting $r_1 = r_2$ in Eq.~\eqref{eq:quantum_kinetic},
\begin{align}
&\partial_t iG_K(r,t;r,t) -\biggl[\epsilon\left(i\vec{\nabla}_{r_1} - \vec{A}(r_1,t)\right)  \nonumber\\
&- \epsilon\left(-i\vec{\nabla}_{r_2} - \vec{A}(r_2,t)\right)\biggr] G_K(r_1,t;r_2,t)\biggr|_{r_1= r_2 = r} = S(rt,rt).\label{kegrad}
\end{align}
Inspecting the second term on the left hand side (lhs) we see that it is a total derivative and the equation can be written as
\begin{align}
&\partial_t n(r,t) - \vec{\nabla}\cdot \vec{j}(r,t) = S(rt,rt),\\
&\vec{j}= \nonumber\\
&\vec{v}\biggl(i\vec{\nabla}_1 - \vec{A}(r_1,t),i\vec{\nabla}_2 + \vec{A}(r_2,t)\biggr)\nonumber
iG_K(r_1,t;r_2,t)\biggr|_{r_1 =r_2 = r},
\end{align}
where the velocity \(\vec{v}\) generalized for an arbitrary dispersion \(\epsilon\), is defined as
\begin{align}
(\vec{x}+\vec{y})\cdot\vec{v}(\vec{x},\vec{y}) = \epsilon(\vec{x}) - \epsilon(-\vec{y}).
\end{align}
This
has the form of a continuity equation for the conserved quantity $n$ because
\begin{align}
S(rt,rt) =0.
\end{align}
We demonstrate this in Appendix~\ref{S1}. Thus the particle number $n(r,t)=iG_K(rt;rt)$ is a conserved quantity.

Conservation of momentum follows similarly by taking the definition of the momentum
\begin{align}
&p^{\alpha}(r,t) = \nonumber\\
&\left[
	\frac{i}{2}
	\left(
		\nabla^\alpha_{r_1} - \nabla^\alpha_{r_2}
	\right)
	-A^\alpha(r,t)
	\right]
iG_K(r_1,t;r_2,t)\biggr|_{r_1 = r_2 = r}.
\end{align}
Evaluating the time-derivative of $p^{\alpha}$, and using
Eq.~\eqref{eq:quantum_kinetic}, one obtains,
\begin{widetext}
\begin{align}
&\partial_t p^\alpha(r,t) - n(r,t)E^\alpha(r,t) + \epsilon^{\alpha\beta\gamma} j^{\beta}(r,t) B^{\gamma}(r,t)- \nabla_r^{\beta}T^{\alpha\beta}(r,t) = 0,\\
&T^{\alpha\beta} =  \biggl[\frac{i}{2}\left(\nabla^\alpha_{r_1}
- \nabla^\alpha_{r_2}\right) -A^\alpha(r,t)\biggr]v^\beta(i\nabla_1 - A(r_1,t),i\nabla_2 + A(r_2,t))iG_K(r_1t,r_2t)\biggr|_{r_1 =r_2 = r} +
\delta^{\alpha\beta}U^{-1}(t)iD_K(rt,rt).
\end{align}
\end{widetext}
Above, the electric field is \(E^\alpha = -\partial_t A^\alpha\) and the magnetic field is
\(B^\alpha = \frac{1}{2}\epsilon^{\alpha\beta\gamma}\left(\nabla^\beta A^\gamma - \nabla^\gamma A^\beta\right)\), and \(T^{\alpha\beta}\)
is the momentum tensor.
Above, we have used the relation
\begin{align}
&\frac{i}{2}\biggl[\nabla_{r_1}^{\alpha} -\nabla_{r_2}^{\alpha}\biggr] S(r_1t,r_2t)\biggl|_{r_1=r_2=r}=\nonumber\\
&=iU^{-1}(t)\nabla^{\alpha}_rD_K(rt,rt).\label{T1eq}
\end{align}
which is proved in Appendix~\ref{T1}. We also note that since \(S(rt,rt)=0\), \(A^{\alpha}(r,t) S(rt,rt)=0\).
Thus, the momentum tensor has a contribution from the full electron Green's function \(G\), and also from the full
superconducting fluctuations \(D\).

Finally conservation of energy is obtained by operating on the kinetic equation $\frac{i}{2}\left(\partial_{t_1} - \partial_{t_2}\right)$
and setting, $r_1 = r_2$, $t_1 = t_2$. This yields,
\begin{widetext}
\begin{align}
&\partial_t \mathcal{E}(r,t) - \vec{\nabla}\cdot \vec{j}_\mathcal{E} =  \vec{E}\cdot \vec{j} +
Q_{\rm ext}(r,t), \\
&\mathcal{E}(r,t) = \frac{i}{2}\left(\partial_{t_1} - \partial_{t_2}\right)iG_K(r,t_1;r,t_2)|_{t_1 = t_2 = t} - {i}U^{-1}(t)D_K(r,t),\\
&j^{\beta}_\mathcal{E} =  \frac{i}{2}\left(\partial_{t_1} - \partial_{t_2}\right)v^\beta\biggl(i\nabla_1 - A(r_1,t_1),
i\nabla_2 + A(r_2,t_2)\biggr)iG_K(r_1,t_1,r_2,t_2)\Big|_{\substack{r_1 =r_2 = r\\t_1 =t_2 =t}},
\\
&Q_{\rm ext} = -{i}D_K(r,t)\partial_t U^{-1}(t).
\end{align}
\end{widetext}
The necessary relation for $S$ needed to derive the above is
\begin{align}
&\frac{i}{2}\left( \partial_{t_1} - \partial_{t_2} \right)S(rt_1,rt_2)|_{t_1 = t_2 = t}\nonumber\\
&=i\partial_t\left( U^{-1} D_K(r,t;r,t) \right) -iD_K(r,t;r,t) \partial_t U^{-1}(t).\label{Sen}
\end{align}
The proof of the above is in Appendix~\ref{aSen}. The above shows that the energy is not conserved due to
Joule heating \(\vec{E}\cdot \vec{j}\), from an applied electric field,
and also when the parameters of the Hamiltonian are explicitly time-dependent \(\partial_t U(t) \neq 0\).

\section{Reduction to a classical kinetic equation}
\label{sec:classical_reduction}

In this section we project the full quantum kinetic equation onto the dynamics of the current. This can be
done by noting some important separation of energy-scales associated with critical slowing down of the
superconducting fluctuations.

\subsection{Critical slowing down}
\label{sec:critical_slowing}
Although the RPA approximation produces a closed set of equations of motion, Eq.~\eqref{eq:RPA_EoM},
these are a set of non-linear coupled integral-differential equations, and therefore cannot be solved without further reduction.
The key to reducing the equations further is the phenomenon of critical slowing down~\cite{HH77}.
This may be demonstrated by considering the particle-particle polarization in equilibrium.
As the equilibrium system is time and space-translation invariant, we consider the Fourier transform
\begin{align}
&\Pi_R(q,\omega) = \int\!d^drdt\,\Pi_R(r,t,r',t')e^{-i\vec{q}\cdot(\vec{r}-\vec{r}') + i\omega(t-t')}
\nonumber\\
&= \frac{i}{2}\int \!d^dk d\eta\, G_R\!\left(k+\frac{q}{2},\eta + \frac{\omega}{2}\right) G_K\!\left(-k+ \frac{q}{2},-\eta +\frac{\omega}{2}\right).
\end{align}
Assuming for the moment that $G_{R}$ has a Fermi-liquid form
\begin{equation}
G_R(k,\omega) = \frac{Z}{\omega - \epsilon_k + i/2\tau} + G_{\rm inc},
\end{equation}
where \(Z\) is the quasiparticle residue,  \(\tau^{-1}\) is the decay rate and is \(\propto {\rm max}(\omega^2,T^2)\), while
\(G_{\rm inc}\) is a smooth incoherent part whose effect is  negligible at low energies. We also assume quasi-equilibrium
where \(G_K\) is related to
\(G_R\) by the fluctuation dissipation theorem \(G_K(k,\omega) = 2{\rm Im}\left[G_R(k,\omega)\right] \tanh(\omega/2T)\).
We expect that $\Pi_R(r,t;r,t')$ decays exponentially on the scale $T^{-1}$. Thus from the perspective of any dynamics that is
slower than $T^{-1}$, the polarization is effectively short ranged in time. Thus, in terms of the Fourier expansion,
we only need the leading behavior in $\omega/T$. Note that the short time dynamics showing the onset of quasi-equilibrium of
the electron distribution function to a temperature \(T\) can also be studied by the numerical time-evolution of the RPA Dyson
equations~\cite{Dasari18}.

Now the Green's function $D_R$ is precisely the linear response to a superconducting fluctuation. Therefore, if $U <U_c$,
where \(U_c\) denotes the critical value of the coupling,
the system is in the disordered phase and $D_R$ should decay exponentially with time. If $U > U_c$ the system is super-critical and
superconducting fluctuations should grow exponentially in time. In terms of the Fourier transform $D_R(q,\omega)$,
this means that there is a pole in the complex $\omega$ plane
which crosses the real axis at precisely the phase transition, $U_c(T)$.  Therefore we may write
\begin{align}
&D_R^{-1}(q,\omega) = U^{-1} - \Pi_R(q,\omega) \nonumber\\
&= Z(q)\biggl[\lambda(q,U) + i\omega/T\biggr] +\mathcal{O}\left(\frac{\omega^2}{T^2}\right),
\end{align}
where the function $\lambda(q = 0,U_c(T)) = 0$.  There is a critical regime in $q$ and $U-U_c$ where
$\lambda(q,U) \ll T$. We further define a thermal wavelength $q_T$ by the expression $\lambda(q_T,U_c) = T$,
which we estimate as $q_T \sim T/v_F$, where \(v_F\) is the Fermi velocity.

On the assumption, to be explicitly shown later,
that the transport is dominated by processes that are much slower than $T^{-1}$, we therefore replace
\begin{equation}
\Pi_R(q;t,t') = U^{-1}(t) - Z(q)\delta(t-t')\biggl[\lambda\left(q, U(t)\right) + \partial_t\biggr],
\end{equation}
where we have explicitly included the possibility that $U(t)$ is time dependent. Therefore the equations of motion for $D$ are,
\begin{align}
\left(\partial_t  + \lambda_q(t)\right)D_R(q,t,t') &= T\nu^{-1}\delta(t-t');\label{Drdef}\\
\lambda_q(t) &= r(t)+ q^2/2M,\label{lamdef}
\end{align}
where \(r(t)\) is an effective detuning arising from the the time-dependence of the interaction \(U(t)\).
Once \(D_R\) is known then, \(D_K\) follows from Eq.~\eqref{Dkdef1}, and the fact that \(\Pi_K(t,t')= 4i\nu \delta (t-t') \).

We consider two cases below. One where
\(\lambda_q(t) =\theta(t) \lambda_q +\theta(-t)r_i\), with \(r_i/T = \mathcal{O}(1)\). This represents a rapid quench, where the
detuning changes from an initial value to a final value at a rate that is faster than or of the order of the temperature.
The second situation we consider is an arbitrary trajectory for the detuning \(r(t)\).

The solution for bosonic propagators for the former, namely the rapid quench, at times \(t,t'\geq 0\) then become,
\begin{align}
\Rightarrow D_R(q,t,t') &= \theta(t-t')T\nu^{-1}e^{-\lambda_q(t-t')},\\
iD_K(q,t,t') &= 2T\nu^{-1}\frac{T}{\lambda_q}\left(e^{-\lambda_q|t-t'|} - e^{-\lambda_q(t+t')}\right),\\
&=\!\! 2\left[D_R(q,t,t')F(q,t') \!+ \!F(q,t)D_A(q,t,t')\right]\label{DKdef},\\
F(q,t)&\equiv\frac{T}{\lambda_q}\left(1 - e^{-2\lambda_qt}\right)\label{Fdefq}.
\end{align}
Note the function $F(q,t)$ changes with rate $\lambda_q$ which for small $q$ is $\ll T$.
Also note that at \(t=t'=0\), the density of superconducting fluctuations as measured by \(D_K(t=t'=0)\)
or \(F(t=0)\) is zero, consistent with an initial condition where the initial detuning is large and
positive, and hence far from the critical point.

For an arbitrary trajectory \(r(t)\) and hence \(\lambda_q(t)\), the solutions for the bosonic propagators, generalize as follows,
\begin{align}
D_R(q,t,t') &= \theta(t-t')T\nu^{-1}e^{-\int_{t'}^t dt''\lambda_q(t'')},\\
iD_K(q,t,t')&=\!\! 2\left[D_R(q,t,t')F(q,t') \!+ \!F(q,t)D_A(q,t,t')\right]\label{DKdef2},\\
F(q,t)&\equiv 2T\int_0^t dx e^{-2\int_x^{t}dy\lambda_q(y)}.
\end{align}
We show in Appendix~\ref{noise} that the above equations for \(D_{R,A,K}\) imply an effective model for a 
bosonic field obeying Langevin dynamics, where the Langevin noise is delta-correlated with a strength proportional to
the temperature \(T\).

\subsection{Linear response to electric field}
\label{sec:linear_response}

The previous derivation of the kinetic equations Eq.~\eqref{eq:quantum_kinetic}
is correct for arbitrary electric fields strengths, and therefore includes non-linear effects.
We now expand the Dyson equation to linear order in $E(t)$ in order to evaluate the linear response.
Note that
since \(g_R(k,t,t')=e^{-i\int_{t'}^t ds\epsilon(k^i+A^i(s))}\), it is the combination,
\(g_R(\vec{k}-\vec{A}(t)) = e^{-i\int_{t'}^t ds\epsilon(k^i+A^i(s)-A^i(t))}\) which is gauge-invariant.
The leading change in the
electron Green's functions at the gauge invariant momentum is,
\begin{align}
g_R(t,t') &= -i\theta(t-t')e^{-i\epsilon_k(t - t')}\nonumber\\
&\times \left(1 -i \int_{t'}^{t} \!\!ds \vec{v}_k \cdot \left[\vec{A}(s)-\vec{A}(t)\right]\right),\label{dEnk}\\
n_k &= n^{\rm eq}_k + \delta n_k(t),
\end{align}
where \(n_k=\int d^d (r_1-r_2) e^{-ik(r_1-r_2)}iG_K(r_1,r_2)\), and
$\delta n$ is assumed to be of order $E$.

We now write the kinetic equation~\eqref{eq:quantum_kinetic} after Fourier transforming with respect to the
spatial coordinates \(r_1-r_2\).
The change in density at the gauge-invariant momentum is
\begin{align}
\frac{d}{dt}n_{k-A(t)}= \partial_t n_k(t) - \partial_t \vec{A}\cdot \vec{\nabla}_k n^{\rm eq}_k = I_C + \vec{E}\cdot \vec{\nabla}_k n^{\rm eq}_k, 
\end{align}
where \(I_C\) is the rhs of the kinetic equation \eqref{eq:quantum_kinetic} evaluated at the gauge-invariant momentum \(\vec{k}-\vec{A}(t)\).
Substituting for \(I_C\), and shifting the internal momentum \(q\) by \(q-2A(t)\), we obtain
\begin{align}
&\partial_t n_k(t) - \vec{E}\cdot \vec{\nabla}_k n^{\rm eq}_k =
	4\sum_q\int\!\!dt' \text{Im}\bigg[\nonumber\\
&	 iD_K(\vec{q}-2\vec{A}(t),t,t')\delta\Pi_A'(\vec{k}-\vec{A}(t),\vec{q}-2\vec{A}(t),t',t)\nonumber\\
	& + i D_R(\vec{q}-2\vec{A}(t),t,t')\delta\Pi_K'(\vec{k}-\vec{A}(t),\vec{q}-2\vec{A}(t),t',t)
	 \nonumber \\
	& +  \delta iD_K(\vec{q}-2\vec{A}(t),t,t')\Pi_A'(\vec{k}-\vec{A}(t),\vec{q}-2\vec{A}(t),t',t) \nonumber\\+
&	 i \delta D_R(\vec{q}-2\vec{A}(t),t,t')\Pi_K'(\vec{k}-\vec{A}(t),\vec{q}-2\vec{A}(t),t,t')
	 \biggl]\label{keqe2},
\end{align}
where the $\delta$ in Eq.~\eqref{keqe2} indicates the first order variation with respect to the electric field.
As expected the conservation laws
continue to hold after this approximation.
Above, the quantity \(\Pi'(k,q,t,t')\) is defined as,
\begin{align}
  &\Pi'_{R,A}(\vec{k}-\vec{A}(t),\vec{q}-2\vec{A}(t),t,t')=\nonumber\\
  &\frac{i}{2} G_{R,A}(\vec{k}-\vec{A}(t),t,t') G_K(-\vec{k}+\vec{q}-\vec{A}(t),t,t'),\\
  &\Pi'_K(\vec{k}-\vec{A}(t),\vec{q}-2\vec{A}(t),t,t')=\nonumber\\
 &\frac{i}{4}\biggl[G_K(\vec{k}-\vec{A}(t),t,t') G_K(-\vec{k}+\vec{q}-\vec{A}(t),t,t')\nonumber\\
&+G_R(\vec{k}-\vec{A}(t),t,t') G_R(-\vec{k}+\vec{q}-\vec{A}(t),t,t')\nonumber\\
&+G_A(\vec{k}-\vec{A}(t),t,t') G_A(-\vec{k}+\vec{q}-\vec{A}(t),t,t')\biggr],\\
&\Pi_{R,A,K}(q,t,t')=\sum_k\Pi_{R,A,K}'(k,q,t,t').
\end{align}
Note that all the momenta appear in the gauge invariant combination \(\vec{k}-\vec{A}(t)\).
Above, the last line highlights the relation between \( \Pi' \), and the polarization \( \Pi \).

The fluctuations are modified by the electric field through their dependence on $\Pi_R$:
\begin{align}
&\delta D_R = D_R\cdot \delta \Pi_R \cdot D_R, \\
&\delta D_K = \delta D_R \cdot \Pi_K \cdot D_A  + D_R \cdot \delta\Pi_K \cdot D_A + D_R \cdot \Pi_K \cdot \delta D_A\nonumber\\
&=  D_R \cdot \delta \Pi_R \cdot D_K  + D_R \cdot \delta\Pi_K \cdot D_A + D_K \cdot \delta \Pi_A \cdot \delta D_A.
\end{align}
In what follows, since we are interested in the response of the current, we multiply the linearized kinetic
equation~\eqref{keqe2} by $v_k$ and integrate over $k$.

\subsection{Large fluctuations limit}
\label{sec:large_fluctuations}

We make one further simplification, which is based on the fact that in the critical regime $D_K \gg D_R$ by a factor $T/\lambda_q$.
Thus we will only keep the terms that are highest order in $D_K$.

\subsection{Projections of equation to finite number of modes}
\label{sec:EOM_projection}

The kinetic equation is an integral-differential equation in which the unknowns $\delta n_k(t)$ appear linearly.
Conceptually therefore it may be solved by standard methods.
However, even though it is linear, it is non-local in time and not time translation invariant.
This makes direct analytical and numerical solution difficult.

We therefore make an additional simplification that instead of considering the full space of solutions,
we instead project entirely onto the current mode. This means that we would like to fix,
\begin{align}
\delta n_k(t)  \overset{?}{=} \frac{\bar{m}}{\rho} J^i(t) \nabla_k^i n^{\rm eq}_k ,\label{eq:wrongJn}\\
\qquad m^{-1}_{ij} = \frac{\partial^2 \epsilon(k)}{\partial k^i \partial k^j}=\partial_{k^j}v_k^i;\,\,
\bar{m}^{-1}\delta_{ij} = -\frac{1}{\rho}\sum_{k} m^{-1}_{ij}n^{\rm eq}_k,
\end{align}
where  \(\rho = -\sum_k n^{\rm eq}_k\) is the density of fermions,
$\bar{m}$ is the effective mass and $J^i(t)$ is some function to be determined. The \(?\) in Eq.~\eqref{eq:wrongJn}
is to question the correctness of this equation. This is because
conservation of momentum
\(P^i=\sum_kk^i n_k\) causes Eq.~\eqref{eq:wrongJn} to 
fail qualitatively. As $\partial_t P^i = \rho E^i $, an electric field pulse, say for simplicity
taken to be a delta-function in
time,  will generate a net momentum. Moreover, by conservation of momentum, the initial perturbation
\begin{equation}
\delta n_k =  V^i k^i \partial n^{\rm eq}/\partial \epsilon,
\end{equation}
with $V^i$ arbitrary, will never decay.
Therefore instead of Eq.~\eqref{eq:wrongJn} we decompose the occupation number as
\begin{align}
\delta n_k(t) &= \frac{\bar{m}}{\rho}
	\left[J_r^i(t)( v_k^i - \gamma k^i/\bar{m})  +  \gamma k^i P^i(t)/\bar{m}^2\right]\partial_\epsilon n^{\rm eq}_k,\\
\gamma &\equiv \frac{\bar{m} \sum_k \vec{v}_k\cdot \vec{k} \partial_\epsilon n}{\sum_k \vec{k}\cdot \vec{k}\partial_\epsilon n},\\
J_{\rm tot}^i &\equiv \sum_k v_k^i \delta n_k = (1-\gamma)J_r^i + \gamma P^i/\bar{m}.\label{Jdef1}
\end{align}
The parameter $\gamma$ gives the amount of current that is carried by the momentum mode,
which does not relax. In the limit $\gamma \rightarrow 0$ there
is no overlap between the modes, the momentum mode carries no current, and is thus neglected. On the other hand,
when $\gamma \rightarrow 1$, as in a Galilean invariant system, the current and momentum are proportional and there is
no relaxing current.

Multiplying the kinetic equation by $v_k^i - \gamma k^i/ \bar{m}$ and $k^i$, and summing over $k$ we obtain,
\begin{align}
&\partial_t J^i_r(t) - \frac{\rho}{\bar{m}}E^i =
	\sum_{k,q}
	\frac{v_k^i -\gamma k^i/m}
		 {1-\gamma}
	4\int\!\!dt' \text{Im}\bigg[\nonumber\\
&\times	 iD_K(q,t,t')\delta\Pi_A'(k,q,t',t)\nonumber\\
	 &+ i D_R(q,t,t')\delta\Pi'_K(k,q, t',t)
	 \nonumber \\
	& +  \delta i D_K(q,t,t')\Pi'_A(k,q,t',t)\nonumber\\
	 &+	 i \delta D_R(q,t,t')\Pi'_K(k,q,t,t')
	 \bigg],\label{Jkin2}\\
&\partial_t P^i(t) - \rho E^i(t) = 0.
\end{align}
Above, in the first equation we have used that $\sum_k (v_k-\gamma k/\bar{m})\delta n_k= (1-\gamma)J^i_r$ as follows
from Eq.~\eqref{Jdef1}. The second equation above just follows from conservation of momentum. We also do not write the
explicit dependence of the momentum labels on the vector potential as it is understood that it is always the gauge-invariant
combination that appears.

As the momentum mode has trivial dependence due to momentum
conservation, we  will simply drop it and set $\gamma = 0$ in Eq.~\eqref{Jkin2}. In this limit,
\begin{eqnarray}
\delta n_k(t) &= \frac{\bar{m}}{\rho}J_r^i(t)\nabla^i_kn_k^{\rm eq}.\label{defn2}
\end{eqnarray}
These are now a closed set of equations in terms of the single unknown function $J_r(t)$ which we will denote simply
as $J(t)$. The remaining step is to evaluate the various terms.
We note that $D_K(t,t')$ is generally larger than $D_R$ by the factor $T/\lambda_q$. Thus in what follows
we will only retain the first and third terms on the right hand side of Eq.~\eqref{Jkin2}.

\subsection{Time dependent kinetic coefficients}
We begin by evaluating the first term, proportional to $\delta\Pi^i_R(q,t,t')$ in Eq.~\eqref{Jkin2}.
There are two contributions, one from varying $n_k$, and we denote it by \(K_J\). The second is from varying $g_R$,
and we denote it by \(K_E\)
\begin{align}
&4\text{Im}\int \!\!dt'\,\sum_{k,q}  v_k^i i D_K(q, t,t')\delta\Pi'_A(k,q,t',t)\nonumber\\
&= K_J^{i}(t) + K_E^{i}(t).
\end{align}
In Appendix~\ref{KJder} we show that the term coming from varying \(n_k\) is
\begin{align}
K_J^i(t) &\simeq \sum_q F_q(t) J^j(t) \tilde{Q}^{ij}_J(t),\nonumber\\
\tilde{Q}_{ij}(t) &\approx
	\frac{\bar{m}\pi q^r q^l }{4\rho}\langle m_{jr}^{-1} m_{il}^{-1} \rangle_{\rm FS}.\label{KJ}
\end{align}
To estimate the magnitude of the above term, we neglect factors of order one, write $\rho \sim \nu k^2_F/m$,
and obtain that this term is $\sim (q/k_F)^2/\nu $.

For the term coming from varying \(g_R\), we show in Appendix~\ref{KEder} that,
\begin{align}
K_E^i(t) &\simeq \sum_q F_q(t) \frac{\rho E^j(t)}{\bar{m}} \tilde{Q}^{ij}_E(t),\nonumber\\
\tilde{Q}^{ij}_E(t) &\approx \frac{28\pi \bar{m}  q^r q^l}{16T \rho}\zeta'(-2)\langle m_{jr}^{-1} m_{il}^{-1} \rangle_{\rm FS}.\label{KE}
\end{align}
The resulting term has the form $\propto (q/k_F)^2 F_q(t) \rho E(t)/\nu T\bar{m}$, and is thus a
parametrically small correction to the drift term
$\rho E/\bar{m}$. Therefore we neglect it for the remainder.

We now turn to the third term $\sim\delta D_K \Pi'_A$ in Eq.~\eqref{Jkin2}.
We begin by considering the change in $\delta D_K$.  This in turn depends on $\delta D_R$,
which from Eq.~\eqref{Drdef} is given by
\begin{equation}
\left[\partial_t + \lambda_q\right]\delta D_R = -\delta \lambda_q D_R.
\end{equation}

There are two contributions to $\delta\lambda_q$. One is from varying $\delta n$ via
\begin{equation}
\lambda_q = U^{-1} -{\rm Re} \biggl[\Pi_R(q)\biggr]=U^{-1} +  \frac{1}{4}\sum_k \frac{n_{k+q/2}+n_{-k+q/2}}{\epsilon_{k+q/2}+\epsilon_{-k+q/2}}.
\end{equation}
Writing the above as a small $q$ expansion, we define $M$ such that,
\begin{align}
\lambda_q = r + \frac{q^2}{2M},
\end{align}
where we find that, on expanding,
\begin{align}
&\sum_k \frac{n_{k+q/2}+n_{-k+q/2}}{2\epsilon_k} = \sum_k\frac{2n_k}{2\epsilon_k} + q^i q^j\sum_k \frac{\nabla^i_k \nabla^j_kn_k}{8\epsilon_k}\nonumber\\
& = \sum_k\frac{2n_k}{2\epsilon_k} + \frac{q^2}{d}\sum_{k}\frac{\nabla^2_k n_k}{8\epsilon_k},
\nonumber \\
&\Rightarrow \frac{1}{M}= \partial_q^2\lambda_q = \frac{1}{4d}\sum_k\frac{\nabla_k^2n_k}{4\epsilon_k}\label{Mdef}.
\end{align}
where $d$ is the spatial dimension. The above equation also shows, $M\sim T/v_F^2$

Thus, the change in \(\lambda_q\) coming from the change in the electron distribution \(\delta n_k\), may be evaluated as follows,
(below we also use $\delta n_k =  -\delta n_{-k}$),
\begin{align}
\delta \lambda_q &=
\frac{1}{4}\sum_k\frac{\delta n_{k+q/2}+\delta n_{-k+q/2}}{\epsilon_{k+q/2}+\epsilon_{-k+q/2}}
\nonumber\\
&= \frac{1}{4}\sum_k\frac{\delta n_{k+q/2}-\delta n_{k-q/2}}{\epsilon_{k+q/2}+\epsilon_{-k+q/2}}\nonumber\\
&\simeq \frac{q^i}{4}\sum_k\frac{\nabla^i_k\delta n_{k}}{2\epsilon_{k}}\qquad (q\rightarrow 0 ).
\end{align}
Using the relation between $\delta n_k$ and the current in Eq.~\eqref{defn2}, and
the expression for $M$ in Eq.~\eqref{Mdef}, we find,
\begin{align}
\delta \lambda_q &= \frac{q^i \bar{m}}{4\rho} J^j\sum_k \frac{\nabla^i_k\nabla^j_k n^{\rm eq}_k}{2\epsilon_k}\nonumber\\
	&=\frac{q^i \bar{m}}{4\rho} J^i\frac{1}{d}\sum_k \frac{\nabla^2_kn^{\rm eq}_k}{2\epsilon_k} \nonumber\\
&=2\frac{\bar{m}}{\rho M}q^i J^i.
\end{align}

The second reason for the change in $\lambda_q$ is due to the direct coupling to the electric field, where defining
$\delta A^i(s) = A^i(s) -A^i(t)$,
\begin{eqnarray}
&&\lambda_q(s) =\biggl[q^i+2\delta A^i(s)\biggr]^2/2M, \nonumber\\
&&\Rightarrow \delta \lambda_q(s) = \frac{2 q^i \delta A^i(s)}{M} = \frac{2q^i}{M}\left[A^i(s)-A^i(t)\right]\nonumber\\
&&=-\frac{2q^i}{M}\int_s^t dt' \partial_{t'} A^i(t'),\nonumber\\
&&\Rightarrow \delta \lambda_q(s)= \frac{2q^i}{M}\int_s^t dt' E^i(t').
\end{eqnarray}
Thus the total change in $\lambda_q$ is
\begin{align}
&\delta\lambda_q(s) = \frac{2\bar{m} q^iJ^i(s)}{ M \rho} +\frac{2q^i}{M}\int_s^t dt' E^i(t')\nonumber\\
&=\frac{2\bar{m} q^i}{ \rho M}\left[J^i(s)+ \int_s^t dt' \frac{\rho E^i(t')}{\bar{m}}\right].\label{dellam1}
\end{align}
\(\delta\lambda_q\) changes both \(D_R\) and \(D_K\), where the change in the former is
\begin{equation}
\delta D_R(t,t' ) = e^{-\lambda_q(t-t')}
	\left[
		-\int_{t'}^t ds \delta \lambda_q(s)
	\right].\label{DR2}
\end{equation}

Note that $\delta D_K = (\delta D_R) \Pi_K D_A+ D_R(\delta \Pi_K) D_A+D_R \Pi_K (\delta D_A)$. 
Since the variation $\delta \Pi_K$ produces a term that is smaller by a factor of $F_q$, we neglect it.
Using the zeroth order calculation that
$i\Pi_K \propto \nu \delta(t -t')$, we have,
\begin{align}
&i\delta D_K(t , t') = \nu \int ds \delta D_R(t,s) D_A(s, t') \nonumber\\
&+ \nu \int ds  D_R(t,s) \delta D_A(s, t').
\end{align}
Substituting in for the result for \(\delta D_R\), using, $\delta D_R(t,t') = \delta D_A(t',t)^*$, we obtain (see Appendix~\ref{Jmemder}
for details),
\begin{align}
&4\int_0^{t} dt' \sum_{k,q}\text{Im}\biggl[v_k^i i \delta D_K(q,t,t') \Pi'_A(k,q,t',t)
\biggr]
 \nonumber\\
 &= \sum_q\alpha q^i q^j \lambda_q
	\int_0^t du
	e^{-2(t-u)\lambda_q} F_q(u)\nonumber\\
&\times	\left[
		J^j(u)+
		\int_u^t dt'
		\frac{\rho E^j(t')}
			 {\bar{m}}
	\right].
\label{Jmem}
\end{align}
The coefficient $\alpha$ is given by,
\begin{align}
\alpha &\equiv \frac{T \bar{m} }{M \nu \rho} \sum_k m^{-1}_{ii}\frac{n^{\rm eq}(\epsilon_k)}{4\epsilon_k^2 + \lambda_q^2}\nonumber\\
&= \frac{ T \bar{m} }{M \nu \rho} \int\!\! d\epsilon\, \kappa(\epsilon)\frac{n^{\rm eq} }{4\epsilon^2 + \lambda^2},\nonumber \\
\kappa(\epsilon) &\equiv  \sum_k m^{-1}_{ii}(k)\delta(\epsilon - \epsilon_k).
\end{align}
The above shows that $\alpha q^2 \sim (q/k_F)^2/\nu$, which is comparable to the local terms. Moreover the sign of $\alpha$ for generic
band structures is
such as to oppose the  local in time term coming from $iD_K\delta\Pi_A$.

The function $\kappa(\epsilon)$ may be Taylor expanded to give $\kappa(\epsilon) \approx  \frac{\nu}{\bar{m}}(1 + b \epsilon/E_F + \cdots)$, where the parameter $b$
is a material dependent parameter of order $1$. The sign of $b$ is not fixed in general.  However for a 'simple' band structure, the parameter $b$ is negative.
Thus,
\begin{align}
\alpha &\approx \frac{ T }{M \rho} \int d\epsilon \left(1+b\frac{\epsilon}{E_F}\right) \frac{n^{\rm eq}(\epsilon) }{4\epsilon^2 + \lambda^2}\nonumber\\
		&= \frac { b T}{M\rho E_F} \int d\epsilon \frac{\epsilon n^{\rm eq}(\epsilon)}{4\epsilon^2}\nonumber\\
&\approx \frac{b T\log(E_F/T)}{M\rho E_F}
\approx \frac{b v^2_F T\log(E_F/T)}{T\nu E_F^2} \nonumber\\
&\approx \frac{1}{k_F^2 \nu} b\log(E_F/T).
\end{align}

For generic quench profiles, the only change that is needed, is the replacement
\begin{align}
e^{-2\lambda_q t}\rightarrow e^{-2\int_0^t ds \lambda_q(s)}.
\end{align}
In what follows we set \(\rho/\bar{m}= 1\). In these units, the ratio \(J/E\) gives a Drude scattering time, and also equals the conductivity.
The kinetic equation for the current for general quench trajectories becomes Eq.~\eqref{jke1}, and for
convenience we rewrite it below,
\begin{align}
&\partial_t J^i(t) - E(t) = -\frac{1}{\tau_r}\biggl[A(t) J^i(t)\nonumber\\
&- \alpha \int_0^t dt' \biggl\{B (t,t') J^i(t') + C(t,t')E^i(t')\biggr\}\biggr].\label{jke1a}
\end{align}
We define the dimensionless quantity,
\begin{align}
A(t) \propto \sum_q q^2 F(q,t).
\end{align}
The memory terms are,
\begin{align}
B(t,s) &\propto \sum_q\left\{ q^2\lambda_q e^{-2\int_s^t\!du\, \lambda_q(u)} F_q(s)\right\},\label{Bdef1}\\
C(t,s) &= \int_0^{s} ds' B(t,s').\label{Cdef}
\end{align}

We now simplify the expressions for \(A,B,C\) using the derived expressions for the time-dependent
superconducting fluctuations.
\begin{align}
A(t)	&= \frac{1}{T^{d/2}}\int_0^t\!ds\, \frac{e^{
	-2 \int_s^t \!du\, r(u)
	}}{(t-s)^{d/2+1} }.\label{Adef}
\end{align}
Similarly the memory term is,
\begin{align}
&B(t,s) \propto \sum_q\left\{ q^2\lambda_q e^{-2\int_s^t\!du\, \lambda_q(u)} F_q(s)\right\}\\
	 &= \frac{1}{T^{d/2}}
	 \int_0^s\!dt'\,e^{-2\int_{t'}^{t}\!du\, r(u)}\nonumber\\	
	&\times  \left(
		 \frac{(1+d/2)+2r(t)(t-t')}{(t-t')^{d/2+2}}
	 \right)
	 .\label{Bdef}
\end{align}

For the case of the sudden quench, since \(r(t) = \theta(t) r + r_i\),
the above expressions for \(A,B,C\) simplify considerably.
For \(r_i/T = \mathcal{O}(1)\) so that
at \(t=0, F(q,t=0)=0\), and for a quench to the critical point \(r=0\), equations \eqref{Adef},
\eqref{Bdef} and \eqref{Cdef} give,
\begin{align}
A_{\rm cr}(t)&= \biggl[1-\frac{2/d}{(1+Tt)^{d/2}}\biggr]\label{Acr}, \\
B_{\rm cr}(t,s)&= T\biggl\{\frac{1}{\left[T(t-s)+1\right]^{1+d/2}}- \frac{1}{\left[Tt+1\right]^{1+d/2}}\biggr\}\label{Bcr},\\
C_{\rm cr}(t,s) &=\frac{2}{d}\biggl\{\frac{1}{\left[T(t-s)+1\right]^{d/2}}- \frac{1+(d/2)\frac{Ts}{(Tt+1)}}{\left[Tt+1\right]^{d/2}}\biggr\}\label{Ccr}.
\end{align}
Above we have regularized the integrals such that a short time cutoff of \(T^{-1}\) has been introduced.
It is also helpful to study the system at non-zero detuning. For a sudden quench to a distance \(r\) from the critical point,
equations ~\eqref{Adef}, ~\eqref{Bdef} and~\eqref{Cdef} give,
\begin{align}
A(r,t)&= 1 - \left(\frac{2r}{T}\right)^{d/2}\Gamma \left(-\frac{d}{2},2 r t\right),\label{Ar}\\
B(r,t,s)&= T\frac{e^{-2r(t-s)}}{(T(t-s))^{1+d/2}} - T\frac{e^{-2rt}}{(Tt)^{1+d/2}},\label{Br}\\
C(r,t,s)&= \left(\frac{2r}{T}\right)^{d/2} \biggl[\Gamma \left(-\frac{d}{2},2 r (t-s)\right)\nonumber\\
&-\Gamma \left(-\frac{d}{2},2 r t\right)\biggr]
-\frac{T s e^{-2rt}}{(Tt)^{1+d/2}}.\label{Cr}
\end{align}

\subsection{Charge Diffusion}

We now adapt the kinetic equation to the case where the current is being driven by a density gradient.
From Eq.~\eqref{eq:quantum_kinetic}, Fourier transforming with respect to the difference in position
coordinates \(r_1-r_2\), and performing a gradient expansion with respect to the center of mass coordinate \(r=\frac{r_1+r_2}{2}\),
\begin{align}
\partial_t n_k(r,t) + \vec{v}_k\cdot\vec{\nabla}n_k(r,t) = S(r,k,t)\label{ch1},
\end{align}
where \(S(k) = \int d^d(r_1-r_2)e^{-ik (r_1-r_2)}S(r_1t,r_2t)\).
A spatial density gradient leads to a current \(J^i =\sum_k v_k^i n_k\). Multiplying Eq.~\eqref{ch1} with
\(v_k\), taking a sum on \(k\), and using the fact that \(v_kS(k)\) in the presence of the current can be written
as a contribution coming from small changes to the superconducting propagator, and small changes to the polarization
bubble as summarized on the rhs of Eq.~\eqref{keqe2}.
The rhs can be simplified as before, leading to the rhs of Eq.~\eqref{jke1a}. This leads to,
\begin{align}
&\partial_t J^i + \sum_k v_k^i v_k^j \nabla^j n_k(r,t)= - \frac{1}{\tau_r}\biggl[A(t)J^i(t)\nonumber\\
&- \alpha
\int_0^tdt' \biggl\{ B(t,t') J^i(t')\nonumber\\
&+ \int dt' C(t,t')E^i(t')\biggr\}\biggr].\label{Diff1}
\end{align}
The coefficients \(A,B\) are given in equations ~\eqref{Adef},~\eqref{Bdef} for a general quench profile, and in
equations ~\eqref{Acr},~\eqref{Bcr} for a rapid quench to the critical point. Although there is no external electric field,
the density gradient induces a current, and the electric field on the rhs is a linear response to this current.

\section{Conductivity and Diffusion coefficient in thermal equilibrium}\label{sec:eqtr}

In thermal equilibrium, one may simply take the long time limit of the expressions in equations~\eqref{Acr},~\eqref{Bcr}, and ~\eqref{Ccr}. Then,
at zero detuning \(r=0\),
\begin{align}
A_{\rm eq}(r=0) &= 1,\nonumber\\
B_{\rm eq}(r=0,t-s) &=\frac{T}{\left[T(t-s)+1\right]^{1+d/2}},\nonumber\\
C_{\rm eq}(r=0,t-s) &=\frac{(2/d)}{\left[T(t-s)+1\right]^{d/2}}.
\end{align}
It is also useful to analyze these coefficients when the system has equilibrated at a non-zero detuning $r$ away from
the critical point. In this case, for $d=2$ we have,
\begin{align}
A_{\rm eq}(r) &= A_{\rm eq}(r=0)\biggl[1+\frac{2r}{T}\ln(2r/T)\biggr],\nonumber\\
B_{\rm eq}(r,t-s) &= T\frac{e^{-2r (t-s)}}{\left[1+ T(t-s)\right]^2} ,\nonumber\\
C_{\rm eq}(r,t-s)\biggl|_{r(t-s) \gg 1} &=\frac{T}{2r}\frac{e^{-2r (t-s)}}{\left[1+ T(t-s)\right]^2}.
\end{align}
Above due to the separation of time-scales discussed in the previous section, \(|r|/T\ll 1\).
We note that in Fourier space,
\begin{align}
\tilde{B}_{\rm eq}(r,\omega=0)&= 1+ \frac{2r}{T}\biggl[\gamma + \ln(2r/T)\biggr],\\
\tilde{C}_{\rm eq}(r,\omega=0)&= \frac{1}{2r}\biggl[1 + \frac{2r}{T}\biggl\{\gamma + \ln(2r/T)\biggr\}\biggr].\label{Ceq}
\end{align}
Above \(\gamma\) is Euler Gamma.

We now discuss the linear response conductivity in thermal equilibrium. In this case, all the coefficients \(A,B,C\) are time-translation
invariant.
Fourier transforming Eq.~\eqref{jke1a}, we obtain,
\begin{align}
J (\omega)= \frac{1+\tau_r^{-1}\alpha \tilde{C}_{\rm eq}(\omega)}
{i\omega +\tau_r^{-1}\biggl(1-\alpha \tilde{B}_{\rm eq}(\omega)\biggr)} E(\omega).\label{lincon}
\end{align}
Writing,
\begin{align}
J = J_{\rm diss} + J_{\rm fl},
\end{align}
where the first is a dissipative current arising due to Drude scattering, while the second is a current arising due to the superconducting
fluctuations. Then, taking the dc limit of Eq.~\eqref{lincon},
\begin{align}
J_{\rm diss} = \tau_r E = \sigma_{\rm diss}E,
\end{align}
while the current from the superconducting fluctuations (neglecting \(B\) which gives small
correction to the Drude scattering rate) is,
\begin{align}
J_{\rm fl} = \alpha \tilde{C}_{\rm eq} E = \sigma_{\rm fl}E.
\end{align}
The above implies that the fluctuation conductivity \(\sigma_{\rm fl}\) gives a correction to the dissipative conductivity \(\tau_r\)
by an amount \(= \tau_r^{-1}\alpha \tilde{C}_{\rm eq}(r,\omega=0)\).
Noting that \(\tau_r\sim T^{-1}\), and using Eq.~\eqref{Ceq}, this implies a fluctuation conductivity correction that goes as
\( \propto \alpha T/r\). While this is qualitatively the same as the fluctuation AL conductivity~\cite{AL68-1,AL68-2},
we note that for the ultra clean case considered here, material dependent parameters such as \(\alpha\) which determine the strength
of Galilean invariance breaking lattice effects, are unavoidable, and give a non-universal material dependent prefactor.

The MT correction is also discussed in the context of a disordered system~\cite{Thompson68,Maki68-1,Maki68-2,Larkin00},
and is well defined only as long as the mean free path is shorter than the inelastic scattering time~\cite{Randeria94,Stepanov18}.
Thus the MT conductivity does not emerge naturally in the clean limit we consider here.

Now we discuss the diffusion constant \(D\) defined
as \(J^i = -D \nabla^i n\). Due to time-translation invariance in equilibrium, we write Eq.~\eqref{Diff1}
in Fourier space. We also note that the electric field generated as a response to the current,
is given by \(J = \tau_r E\), up to fluctuation corrections.
This gives the diffusion constant,
\begin{align}
D(\omega) = \frac{v_F^2/2}{i\omega +  \tau_r^{-1}
\biggl\{1-\alpha\biggl(\tilde{B}_{\rm eq}(\omega)+ \frac{\tilde{C}_{\rm eq}(\omega)}{\tau_r}\biggr)\biggr\}}.
\end{align}
Close to the critical point, \(C\) dominates over \(B\). Moreover, Taylor expanding in \(\alpha\),
the dc limit of the diffusion constant is,
\begin{align}
D(\omega=0) \approx \frac{v_F^2\tau_r}{2}\biggl[1 + \frac{\alpha}{2\tau_rr}\biggr].
\end{align}
Thus the fluctuation correction for the diffusion constant has qualitatively the same form as that for
the conductivity in being \(\propto \alpha T/r\), where we have used that \(\tau_r \sim T^{-1}\).

\section{Solving the kinetic equation for the conductivity}
\label{sec:solving}

The solution of the kinetic equation for the current was presented in much detail
in Ref.~\onlinecite{Lemonik17b}. For completeness we summarize some
of the findings. Two kinds of quench trajectories were studied.
One was a rapid quench from deep in the disordered phase to a distance \(r\geq 0\) away from the
critical point. The second was a smooth quench trajectory which started from the disordered phase, approached the
critical point \(r=0\), and returned back to the disordered phase. Regardless of the details of the trajectory,
the current showed slow dynamics because of slowly relaxing superconducting fluctuations. In the kinetic equation Eq.~\eqref{jke1a},
this physics is encoded in the memory terms ($B,C$).

In addition, the dynamics of the current for a critical quench to \(r=0\), was found to show universal
behavior, with a power-law aging in the conductivity \(\sigma(t,t')\), a result also found for a disordered system~\cite{Lemonik18}.
Moreover, the dynamics at non-zero detuning \(r\) was shown to obey scaling collapse. However it should be noted that, for
the clean system studied here, the exponents
entering the scaling behavior were non-universal in that they depend on~\cite{Lemonik17b} \(T\tau_r\).

For a smooth quench it was shown that
transiently enhanced superconducting fluctuations create transient a low resistance current carrying channel, whose signature 
is a suppression of the  Drude scattering rate at low frequencies. Note that  
the Drude scattering rate in the absence of time-translation invariance is defined as,
\begin{align}
\tau_{\rm Dr}(t,\omega) = -\frac{{\rm Im}\left[\sigma(\omega,t)\right]}{\omega {\rm Re}\left[\sigma(\omega,t)\right]},
\end{align}
with
\begin{align}
\sigma(\omega,t) =\int d\tau e^{i\omega \tau}\sigma(t+\tau,t).
\end{align}

\begin{figure}[H]
  \begin{subfigure}[t]{0.49\textwidth}
  \includegraphics[width = \textwidth]{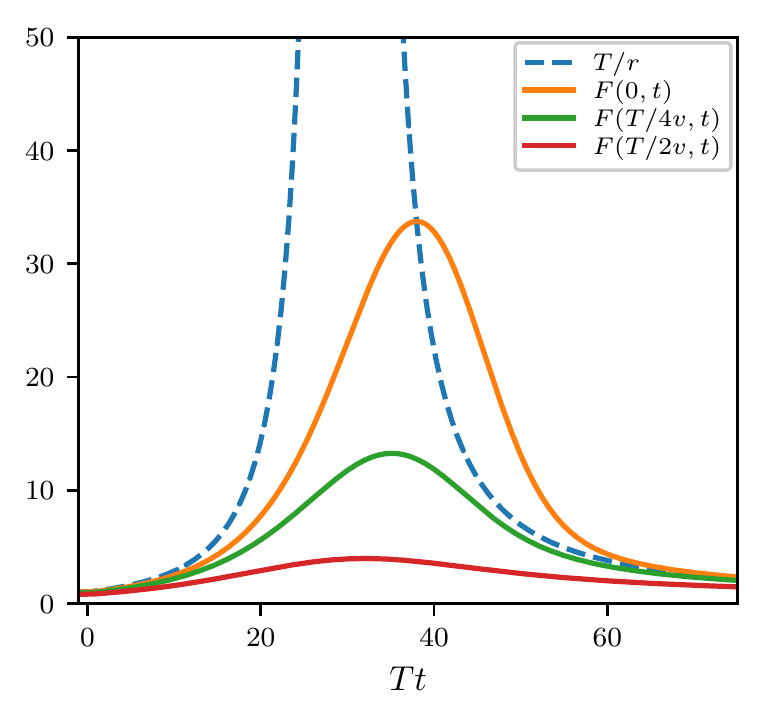}
\caption{Fluctuations \(F(q,t)\) for three different \(q\) and for a critical quench where the detuning varies smoothly as $r(t)/T = 1.0 -\theta(t)(Tt/30)e^{1-Tt/30}$.
In thermal equilibrium \(F(q=0)=T/r\). As a reference, \(T/r(t)\) is plotted.}
  \end{subfigure}
  \begin{subfigure}[t]{0.49\textwidth}
  \includegraphics[width = \textwidth]{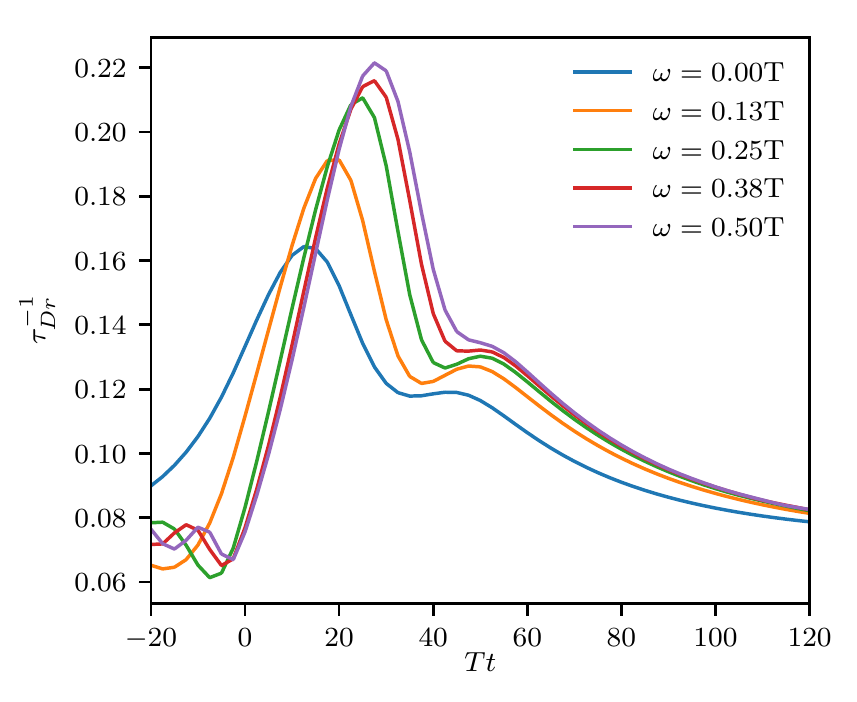}
  \caption{Drude scattering rate for the critical quench $r(t)/T = 1.0 -\theta(t)(Tt/30)e^{1-Tt/30}$.
The high frequency Drude scattering
rate follows the profile of the fluctuations (top figure) by smoothly increasing and decreasing
in time. On the other hand, the low frequency Drude scattering rate first increases, followed by being suppressed at approximately
when the superconducting fluctuations peak.}
  \end{subfigure}
\caption{Critical quench}\label{crsm}
\end{figure}

\begin{figure}[H]
  \begin{subfigure}[t]{0.49\textwidth}
  \includegraphics[width = \textwidth]{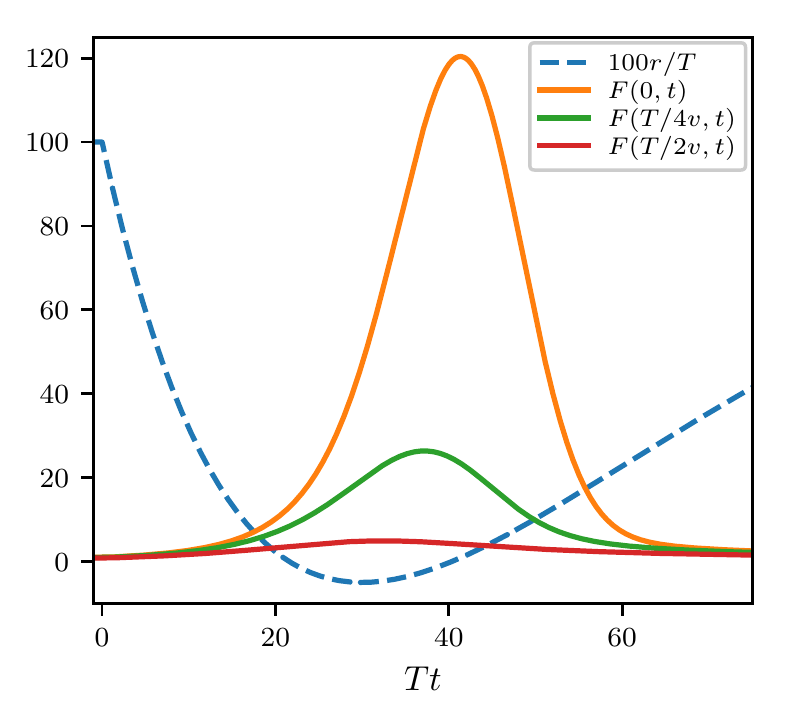}
\caption{Fluctuations \(F(q,t)\) for three different \(q\) and for a super-critical quench where the detuning varies smoothly
as $r(t)/T = 1.0 -\theta(t)(1+0.05)(Tt/30)e^{1-Tt/30}$, becoming negative for a certain length of time.
In thermal equilibrium \(F(q=0)=T/r\). As a reference, \(100*r(t)/T\) is plotted.}
  \end{subfigure}
  \begin{subfigure}[t]{0.49\textwidth}
  \includegraphics[width = \textwidth]{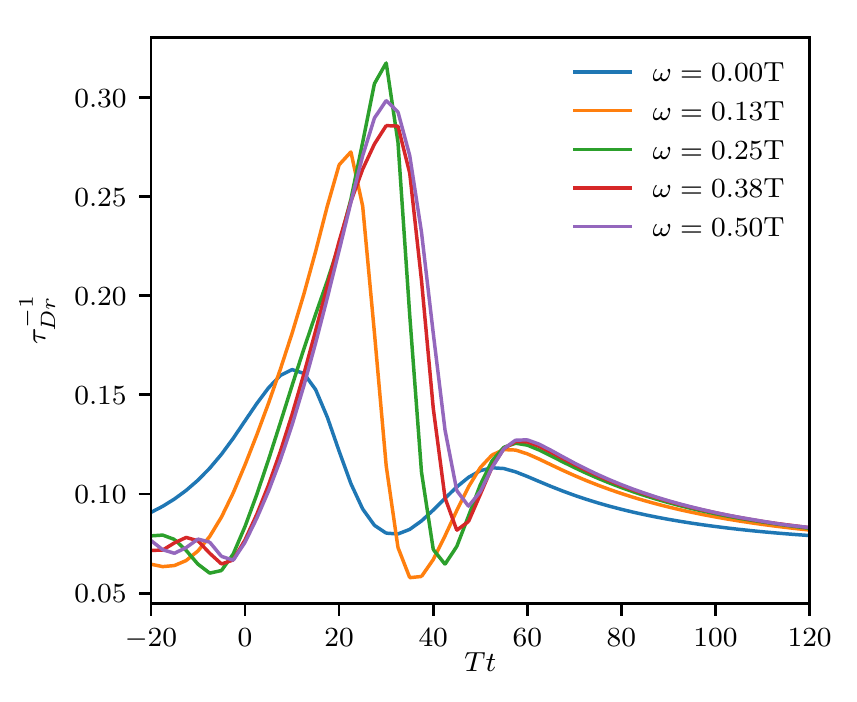}
  \caption{Drude conductivity for a super-critical quench where $r(t)/T = 1.0 -\theta(t)*(1+0.05)(Tt/30)e^{1-Tt/30}$. Qualitatively similar behavior
as in Fig.~\ref{crsm}, but all the effects are more amplified. As
the density of superconducting fluctuations \(F(q,t)\) (top figure) grow in time, the high frequency
Drude scattering  rate increases, while the low frequency Drude scattering rate is suppressed at approximately when the superconducting
fluctuations peak. The behavior in frequency and time is also more dispersive for the super-critical quench than the critical quench.}
  \end{subfigure}
\caption{Super-critical quench}\label{scrsm}
\end{figure}

In this section we revisit the smooth quench, defined by the trajectory
\begin{align}
r(t)/T = 1 - \theta(t)\left(1+\epsilon\right)(t T/30) e^{1 - t T/30}.
\end{align}
Above, \(r/T\) starts out being \(1\), smoothly approaches \(-\epsilon\) at a time \(T t_* = 30\), and then
smoothly returns back to its initial value of \(r/T=1\). We consider two cases, one where \(\epsilon=0\) (see Fig.~\ref{crsm}),
and one where the quench is  super-critical in that \(\epsilon =0.05\). For the latter case, for a finite time, the parameters of the Hamiltonian correspond
to that of an ordered phase since \(r<0\) (see Fig.~\ref{scrsm}). Despite this, for such a transient regime, the order-parameter is still zero,
although superconducting fluctuations are significantly more enhanced than if \(r\geq 0\). In particular, critical slowing down
prevents true long range order to
develop over time scales over which the microscopic parameters are varying.

Note that in the numerical simulation, we apply an electric field pulse which is a delta-function in time. In particular,
for an electric field pulse of
unit strength centered at \(t'\), the conductivity equals the current, \(\sigma(t,t')= J(t)\).
Fourier transforming with respect to
\(t-t'\), we can obtain the conductivity
for arbitrary \(\omega\). However, in an actual experiment, the finite temporal width of the electric field pulse places a limit on the
lowest frequency accessible. Nevertheless, the physics of a suppressed Drude scattering at low frequencies is
visible over a sufficiently broad range of frequencies, as shown in figures \ref{crsm} and \ref{scrsm}, for this to be a feasible experimental
observation.

Top panels of Fig.~\ref{crsm} and Fig.~\ref{scrsm} show how the superconducting fluctuations, at several different wavelengths,
evolve for a critical and a super-critical quench
respectively. The lower panels of the same figures show how the corresponding Drude scattering rate, for different frequencies, evolve in time.
When the density of superconducting fluctuations peak, the Drude scattering rate at low frequencies \(\omega <T\) dips, with the effect being more
enhanced for the super-critical quench. Moreover the Drude scattering rate is strongly dispersive in that different frequency
components of the Drude scattering rate peak at different times.

Recall that when the system returns to the normal phase, since we are in the clean limit, the steady-state
Drude scattering rate is zero. A slow power-law approach to thermal equilibrium is also
visible in the long time tails of the Drude scattering rate that are found to persist long after the detuning has returned back to its initial
value in the normal phase. The slow dynamics highlighted above provide signatures in time-resolved optical conductivity where, although the system
lives for too short a time for true long range order to develop, the transient dynamics can still show clear signatures of superconducting fluctuations.

\section{Spectral properties}
\label{sec:spectral}
We now present results for the time-evolution of the electron spectral properties. Results for
the electron life-time obtained from the imaginary part of the electron self-energy, appear Ref.~\onlinecite{Lemonik17}, where it was
shown that the behavior is non-Fermi liquid like
due to enhanced Andreev scattering of the electrons at the Fermi energy. In this section we discuss how this
scattering affects the local density of states.

For simplicity we
consider a rapid quench from deep in the normal phase where \(F(q)=0\) initially, to a final
arbitrarily small detuning \(r\geq 0\) away from the critical point.
For this quench, the superconducting fluctuations obey the equation,
\begin{equation}
\left[ \frac{1}{T} \partial_t + 2\left(\frac{v^2}{T^2}q^2 + r \right) \right] F(q^2,t)  = 2,\label{Feq1}
\end{equation}
where the solution of the above is identical to Eq.~\eqref{Fdefq} with \(\lambda_q/T = r + v^2 q^2/T^2\). We have assumed that
$F$ depends only isotropically on $q$, and follow a convention where \(r\) is dimensionless. We define the dimensionless
variable $x=v^2q^2/T^2$. The electron self energy using Eq.~\eqref{Sr} and in the limit of $D_K\gg D_R$, and in two spatial dimensions is,
\begin{align}
\Sigma_{R,A}(k,t_1,t_2)&= \frac{i}{2}\int \frac{d^2q}{(2\pi)^2} D_K(q,t_1,t_2)\nonumber\\
&\times G_{A,R}(-k+q,t_2,t_1).
\end{align}
Going into Wigner coordinates which involves Fourier transforming with respect to
relative coordinates $t_1-t_2$, while keeping the mean time $t=(t_1+t_2)/2$ dependence in $D_K$~\cite{Lemonik17},
and performing an angular integral, we obtain,
\begin{align}
&\frac{\Sigma_R\left(\omega,\varepsilon\right)}{T} = + \frac{\text{Gi}}{4\pi} \int_0^\infty dx F(x,t)\nonumber\\
&\times \left[\left(\frac{\omega + \varepsilon + \Sigma_A\left(-\omega,\varepsilon \right)}{T}\right)^2 - x \right]^{-1/2}, \label{SR}\\
&\frac{\Sigma_A\left(-\omega,\varepsilon\right)}{T} = + \frac{\text{Gi}}{4\pi} \int_0^\infty dx F(x,t)\nonumber\\
&\times \left[\left(\frac{-\omega + \varepsilon + \Sigma_R\left(\omega,\varepsilon \right)}{T}\right)^2 - x \right]^{-1/2}, \label{SA}\\
&\text{where} \,\,\, \text{Gi} \equiv  \frac{8T}{\pi\nu v^2}.
\end{align}

We start in the perturbative limit by setting $\Sigma_A = i0^+$, $\Sigma_R = -i0^+$. First let us study
the equilibrium case. Here $F = 1/(x + r)$. Thus we have,
\begin{align}
&\frac{\Sigma_R\left(\omega,\varepsilon\right)}{T} =   \frac{\text{Gi}}{8\pi} \int_0^\infty \frac{dx}{x + r} \nonumber\\
&\times \left[\left(\frac{\omega + \varepsilon + \Sigma_A\left(-\omega,\varepsilon \right)}{T}\right)^2 - x \right]^{-1/2}.
\\
&\qquad\text{Define } \bar{x} \equiv \frac{ x}{r}; \quad z_+^0 \equiv \frac{\omega + \varepsilon +i0^+}{T}\sqrt{\frac{1}{r}},
\\
&\frac{\Sigma_R\left(\omega,\varepsilon\right)}{T}
=\frac{\text{Gi}}{8\pi} \sqrt{\frac{1}{r}} \int_0^\infty\frac{d\bar{x}}{1+\bar{x}}\left[\left(z_+^0\right)^2 - \bar{x} \right]^{-1/2}
\nonumber\\
& = \frac{\text{Gi}}{8\pi\sqrt{r}} f(z_0^+),
\\
&\text{where} \,\,\,\text{Re}\biggl[f(z)\biggr] \equiv \frac{1}{\sqrt{z^2 + 1}}\log\frac{z + \sqrt{1+z^2}}{-z + \sqrt{1+z^2}}.\label{fzdef}
\end{align}
The density of states is,
\begin{equation}
\nu(\omega) = -\frac{\nu_0}{\pi} \int d\varepsilon \Im \left[G^R\left(\varepsilon,\omega\right)\right].
\end{equation}
To evaluate this integral, we need to locate the pole in $\varepsilon$ complex plane:
$\omega - \varepsilon^*(\omega) -\Sigma^R(\varepsilon^*(\omega),\omega) = 0$.
Then the value of the integral follows from the residue
\begin{equation}
\nu(\omega) = \frac{\nu_0}{\pi} \Im \biggl[\frac{ \pi i}{1 + \frac{\partial \Sigma^R}{\partial\varepsilon}|_{\varepsilon =\varepsilon^*}}\biggr].
\label{nuch}
\end{equation}

Thus, the perturbative change in the density of states is obtained from Taylor expanding,
\begin{align}
\frac{\delta \nu(\omega)}{\nu} &= -\text{Re} \biggl[\partial_\varepsilon \Sigma_R\left(\omega,\varepsilon\right)\biggr]\biggl |_{\varepsilon = \omega}
\nonumber\\
& = -\text{Re} \sqrt{\frac{1}{r}}\frac{\text{Gi}}{8\pi}\sqrt{\frac{1}{r}} \partial_z f(z)|_{z = \frac{2\omega}{T\sqrt{r}}} \nonumber\\
& = \frac{\text{Gi}}{8\pi r}  g\left(\frac{2\omega}{T\sqrt{r}} \right), \\
&\text{where}\qquad g(z) = -\text{Re} \biggl[\partial_z f(z)\biggr].
\end{align}
Using Eq.~\eqref{fzdef}, one obtains the following asymptotic behavior,
\begin{equation}
\frac{\delta \nu (\omega) }{\nu} = \frac{\text{Gi}}{4\pi r}
\begin{cases}
-1 & \omega \ll T\sqrt{r} \\
\frac{ r T^2}{8 \omega^2} \log\left(\frac{4\omega^2 }{r T^2}\right) & \omega \gg T\sqrt{r}
\end{cases}
\end{equation}
Note that this shift is perturbative for all $\omega$, that is  $\delta\nu/\nu\ll 1$, when $\text{Gi}/ 4\pi r \ll 1$.

Now we turn to the critical quench, where $ r$ is $\mathcal{O}(1)$ for $ t<0$ and $r = 0$ for $t > 0$. In this case, the
solution of Eq.~\eqref{Feq1} gives,
\begin{equation}
F(x,t) = \frac{1 - e^{-2 x Tt} }{x}.
\end{equation}
The above implies,
\begin{align}
&\frac{\Sigma_R\left(\omega,\varepsilon\right)}{T}
 =
 \frac{\text{Gi}}{8\pi} \int_0^\infty dx\frac{1-e^{-2xTt}}{x}\nonumber\\
&\times \left[\left(\frac{\omega + \varepsilon + i0^+}{T}\right)^2 - x \right]^{-1/2},
\\
&\qquad\text{define } \bar{x} = 2 x T t; \quad z_+^0 \equiv \frac{\omega + \varepsilon +i0^+}{T}\sqrt{2tT},
\\
&\frac{\Sigma_R\left(\omega,\varepsilon\right)}{T}
=\frac{\text{Gi}}{4\pi} \sqrt{\frac{Tt}{2}} \int_0^\infty\! d\bar{x}\,\frac{1- e^{-\bar{x}}}{\bar{x}}\left[\left(z_+^0\right)^2 - \bar{x} \right]^{-1/2}
\nonumber\\
& = \frac{\text{Gi}}{8\pi}\left(\sqrt{2Tt}\right) h(z_0^+).
\end{align}
In Appendix~\ref{hderapp} we show that \(h\) has the following asymptotic form,
\begin{equation}
\text{Re}\left[h(z)\right] =
\begin{cases}
2z & |z| \ll 1 \\
z^{-1}\biggl(\log\left(z^2\right) + \Gamma\left(\frac{1}{2}\right)\biggr) & |z| \gg 1
\end{cases}\label{hder}
\end{equation}

Proceeding to the change in the density of states we have
\begin{align}
\frac{\delta \nu \left(\omega\right) }{\nu}
&=
-\frac{\text{Gi} } {4\pi \sqrt{2} } \sqrt{Tt} \, \text{Re} \left[T\partial_\varepsilon h\left(z_+^0\right)\right]|_{\varepsilon = \omega} \nonumber \\
& = -\frac{\text{Gi} } {8\pi } (2Tt) \,  \text{Re}\left[\partial_z h\left(z\right)\right]|_{z = 2\sqrt{2Tt}\frac{\omega}{T}} \\
& =  \frac{\text{Gi}}{4\pi}
\begin{cases}
-2Tt & \omega \ll \sqrt{T/t} \\
\frac{1}{8}\frac{T^2}{\omega^2} \log\left(\omega^2 t/T\right) & \omega \gg \sqrt{T/t}
\end{cases}
\end{align}

Note that the asymptotic behaviors of the critical and equilibrium cases agree if we set $r = \frac{1}{2 T t}$. This follows from the fact that
the fluctuations $F(x,t)$ behave in a comparable manner in these two cases.

\subsection{Self-Consistent Regime (Results for $\omega=0$)}

We now turn to the full solution that does not require \(\delta \nu \ll \nu\).
We will first work out the equilibrium case. The critical quench case will then follow from the substitution of $r = \frac{1}{2 T t}$.

By adding $\pm \omega + \varepsilon$, the self-consistent equation for the
self-energy in equations ~\eqref{SR},~\eqref{SA} may be written as,
\begin{gather}
z_+ =  z_+^0 + \gamma f\left(z_-\right),\label{SRa}\\
z_- =  z_-^0 + \gamma f\left(z_+\right),\label{SAa}
\end{gather}
where
\begin{gather}
z_+ \equiv \frac{\omega +\varepsilon + \Sigma_A\left(-\omega,\varepsilon\right)}{T}\sqrt{\frac{1}{r}},\\
z_- \equiv \frac{- \omega +\varepsilon + \Sigma_R\left(\omega,\varepsilon\right)}{T} \sqrt{\frac{1}{r}},\\
\gamma = \frac{\text{Gi}}{8\pi r}.
\end{gather}
The above self-consistent equations can be solved in certain limits (see Appendix~\ref{nuder}) giving the following
results for the density of states at zero frequency,
\begin{equation}
\nu(\omega = 0) \propto \frac{\nu_0}{\log \left(\text{Gi}/8\pi r\right)}.\label{selfnu}
\end{equation}
The critical quench case then follows by the substitution $r = \frac{1}{2 T t}$,
\begin{equation}
\nu(\omega = 0,t) \propto \frac{\nu_0}{\log \left(\text{Gi}T  t/4\pi\right)}.\label{selfnu2}
\end{equation}
Thus we find that there is no true steady state of the zero frequency density of states due to the solution continuing to evolve
logarithmically in time. This time-evolution, although slow, happens
because there is no non-zero temperature critical point in \(d=2\). To obtain the correct thermalized steady-state,
vortices need to be accounted for in our treatment. However, for the transient regime relevant for experiments, where no
superconducting gap has developed yet, and the dynamics is governed by weakly interacting superconducting
fluctuations, our results are still valid.

\section{Conclusions} \label{sec:Concl}
We have studied quench dynamics of an interacting electron system along general quench trajectories
using a 2PI approach. To leading non-trivial order in $1/N$, our treatment reduced to RPA in the
particle-particle channel. While the quantum kinetic equations are exact, leading to proper definitions of
the conserved densities and currents, progress was made in solving them by exploiting a separation of time-scales.
This could be seen by noting that
thermalization of the electron distribution function sets in on a time-scale of \(T^{-1}\), while
the collective modes relax at a rate
given by the detuning to the superconducting critical point, which can be arbitrarily slow. This is the phenomena of
critical slowing down, and we derived an effective classical equation for the decay of the current in this regime.
The dynamics of the current was also justified using a phenomenological mapping to model F in the Halperin-Hohenberg
classification scheme. Going forward, we envision that such mappings may be helpful in other contexts when understanding
time-resolved experiments that aim to probe collective modes.

Results were also presented for the local density of states which was found to be enhanced at low frequencies due to
Andreev scattering processes. For a critical quench, the self-consistent equations showed an equivalence between
the time after the quench and the inverse detuning.

Future directions involve studying the coarsening regime, in particular how true long range order develops. The reverse quench
where the initial state is ordered, also needs to be explored, going beyond mean-field.
Experiments involving ultra-fast manipulation of spin and charge order also require generalizing our study to
quench dynamics in the particle-hole channel.

Acknowledgments: This work was supported by the
US National Science Foundation Grant NSF-DMR 1607059.

\begin{appendix}

\section{Proving conservation laws}

\subsection{Proof of \(S(rt,rt)=0\)} \label{S1}

Using Eq.~\eqref{SigDef},
\begin{align}
S(rt,rt) &=
\frac{i}{2}\int dy \biggl\{  G_A(y;r,t)G_A(y;r,t)D_R(r,t;y) \nonumber\\
&+  G_K(y;r,t)G_A(y;r,t)D_K(r,t;y) \biggr\} \nonumber\\
&+\frac{i}{2}\int dy \biggl\{G_K(y;r,t)G_K(y;r,t)D_R(r,t;y)\nonumber\\
&+ G_A(y;r,t)G_K(y;r,t)D_K(r,t;y)\biggr\} \nonumber\\
&-\frac{i}{2}\int dy \biggl\{G_R(r,t;y)G_R(r,t;y)D_A(y;r,t) \nonumber\\
&+  G_R(r,t;y)G_K(r,t;y)D_K(y;r,t)\biggr\} \nonumber\\
&-\frac{i}{2}\int dy \biggl\{G_K(r,t;y) G_R(r,t;y)D_K(y;r,t)\nonumber\\
&+ G_K(r,t;y)G_K(r,t;y)D_A(y;r,t)\biggr\}.\label{qkef}
\end{align}
The above terms can be rearranged to give,
\begin{eqnarray}
&S(rt,rt)=
2\int dy \biggl\{\Pi_K(y;r,t)D_R(r,t;y) \nonumber\\
&+ \Pi_A(y;r,t)D_K(r,t;y) \biggr\} \nonumber\\
&- 2\int dy \biggl\{\Pi_K(r,t;y)D_A(y;r,t) \nonumber\\
&+  \Pi_R(r,t;y)D_K(y;r,t)\biggr\}.
\end{eqnarray}
Using Eq.~\eqref{Dkdef1},
the second and fourth terms above are rewritten to give,
\begin{eqnarray}
&&S(rt,rt)=
2\int dy \biggl\{D_R(r,t;y)\Pi_K(y;r,t) \nonumber\\
&&+ \biggl(D_R \Pi_K D_A\biggr)_{r,t;y}\Pi_A(y;r,t) \biggr\} \nonumber\\
&&- 2\int dy\biggl\{\Pi_K(r,t;y)D_A(y;r,t) \nonumber\\
&&+  \Pi_R(r,t;y)\biggl(D_R \Pi_K D_A\biggr)_{y;r,t}\biggr\}.
\end{eqnarray}
Using Eq.~\eqref{Drdef1} and writing, $\Pi_{A,R}D_{A,R} = U^{-1}D_{A,R}-1$,  or
equivalently $D_{A,R}\Pi_{A,R} = D_{A,R}U^{-1}-1$,
\begin{eqnarray}
&&S(rt,rt)
=2U^{-1}(t)\biggl[ D_R \Pi_K D_A\biggr]_{r,t;r,t} \nonumber\\
&&- 2U^{-1}(t){\rm Tr}\biggl[D_R\Pi_K D_A\biggr]_{r,t;r,t} =0.
\end{eqnarray}

\subsection{Proof of Eq.~\eqref{T1eq}}\label{T1}

It is convenient to write $S$ for unequal positions and times, and express the
self-energy \(\Sigma\) in terms of the \(G,D\) propagators.
Doing this we obtain,
\begin{align}
&S(r_1t_1,r_2t_2)= \frac{1}{2}\int_{y}\biggl[i\biggl\{D_K(r_1t_1,y)G_K(y,r_1t_1) \nonumber\\
&+ D_R(r_1t_1,y) G_A(y,r_1t_1)\nonumber\\
&+ D_A(r_1t_1,y)G_R(y,r_1t_1)\biggr\} G_A(y,r_2t_2) \nonumber\\
&+ i\biggl\{D_R(r_1t_1,y) G_K(y,r_1t_1)\nonumber\\
&+ D_K(r_1t_1,y) G_A(y,r_1t_1)\biggr\}G_K(y,r_2t_2)\biggr]\nonumber\\
&-\frac{1}{2}\int_{y}\biggl[iG_R(r_1t_1,y)\biggl\{D_K(y,r_2t_2)G_K(r_2t_2,y) \nonumber\\
&+ D_R(y,r_2t_2) G_A(r_2t_2,y)+ D_A(y,r_2t_2)G_R(r_2t_2,y)\biggr\} \nonumber\\
&+ iG_K(r_1t_1,y)\biggl\{D_A(y,r_2t_2)G_K(r_2t_2,y) \nonumber\\
&+ D_K(y,r_2t_2) G_R(r_2t_2,y)\biggr\}\biggr].\label{Sgen}
\end{align}

In order to prove momentum conservation we need to study the action of the spatial gradients on \(S\),
\begin{align}
&\frac{i}{2}\biggl[\nabla_{r_1}^{\alpha} -\nabla_{r_2}^{\alpha}\biggr] S(r_1t_1,r_2t_2)\biggl|_{1=2}=\nonumber\\
& \int_{r't'}\biggl[\biggl\{i\nabla_{r_1}^{\alpha}D_K(r_1t_1,r't')\biggr\}\Pi^A(r't',r_1t_1)\nonumber\\
&+ \biggl\{i\nabla_{r_1}^{\alpha}D_R(r_1t_1,r't')\biggr\}\Pi^K(r't',r_1t_1)\nonumber\\
&+ \int_{r't'}\biggl[\Pi^R(r_1t_1,r't')\biggl\{i\nabla_{r_1}^{\alpha}D_K(r't',r_1t_1)\biggr\} \nonumber\\
&+ \Pi^K(r_1t_1,r't')\biggl\{i\nabla_{r_1}^{\alpha}D_A(r't',r_1t_1)\biggr\}\biggr].\label{Sint2}
\end{align}
where we have used that $\left(\partial_{r_1}-\partial_{r_2}\right) \left[G_{a}(1',1)G_b(1',2)+G_b(1',1)G_a(1',2)\right]\biggl |_{1=2}=0$.
Now using that
\begin{align}
&\int_{r't'}\biggl\{i\nabla_{r_1}^{\alpha}D_K(r_1t_1,r't')\biggr\}\Pi^A(r't',r_1t_1)= \nonumber\\
&\int_{r't',.,..}i\nabla_{r_1}^{\alpha}D_R(r_1t_1,.)\Pi_K(.,..)D_A(..,r't')\Pi^A(r't',r_2t_1)\biggl|_{r_2=r_1}\nonumber\\
&= \int_{r't',.,..}i\nabla_{r_1}^{\alpha}D_R(r_1t_1,.)\Pi_K(.,..)\biggl[D_A(..,r_2t_1)U^{-1}(t_1)\biggl|_{r_2=r_1}\nonumber\\
&-\delta_{..,1}\biggr]\nonumber\\
&= i\nabla_{r_1}^{\alpha}D_K(r_1t_1,r_2t_1)U^{-1}(t_1)\biggl|_{r_1=r_2} \nonumber\\
&- \int_{r't'}\biggl\{i\nabla_{r_1}^{\alpha}D_R(r_1t_1,r't')\biggr\}\Pi_K(r't',r_1t_1)\nonumber,
\end{align}
and performing a similar analysis for second last term in Eq.~\eqref{Sint2}, we obtain,
\begin{align}
&\frac{i}{2}\biggl[\nabla_{r_1}^{\alpha} -\nabla_{r_2}^{\alpha}\biggr] S(r_1t,r_2t)\biggl|_{r_1=r_2=r}=\nonumber\\
&= {i}U^{-1}(t)\nabla^{\alpha}_{r_1}D_K(r_1t,r_2t)\biggr|_{r_1=r_2=r}\nonumber\\
&+ {i}U^{-1}(t)\nabla^{\alpha}_{r_2}D_K(r_1t,r_2t)\biggr|_{r_1=r_2=r}\nonumber\\
&=iU^{-1}(t)\nabla^{\alpha}_rD_K(rt,rt).
\end{align}

\subsection{Proof of Eq.~\eqref{Sen}}\label{aSen}

The manipulations involved are,
\begin{align}
&\frac{i}{2}\left( \partial_{t_1} - \partial_{t_2} \right)S(rt_1,rt_2)|_{t_1 = t_2 = t}
\nonumber\\
&=\int_{r't'}\biggl[\biggl\{i\partial_{t_1}D_K(r_1t_1,r't')\biggr\}\Pi^A(r't',r_1t_1)\nonumber\\
&+ \biggl\{i\partial_{t_1}D_R(r_1t_1,r't')\biggr\}\Pi^K(r't',r_1t_1)\nonumber\\
&+\int_{r't'}\biggl[\Pi^R(r_1t_1,r't')\biggl\{i\partial_{t_1}D_K(r't',r_1t_1)\biggr\} \nonumber\\
&+ \Pi^K(r_1t_1,r't')\biggl\{i\partial_{t_1}D_A(r't',r_1t_1)\biggr\}\biggr]
\nonumber\\
=&i\partial_{t_1} D_K(r,t_1;r,t_2) U^{-1}(t_2)|_{t_1 = t_2 = t} \nonumber\\
&+ i\partial_{t_2}U^{-1}(t_1) D_K(r,t_1;r,t_2) |_{t_1 = t_2 = t}
\nonumber\\
= &iU^{-1}(t)\left[\partial_{t_1} D_K(r,t_1;r,t_2)+\partial_{t_2} D_K(r,t_1;r,t_2) \right]|_{t_1 = t_2 = t}
\nonumber\\
= &iU^{-1}\partial_t D_K(r,t;r,t)
\nonumber\\
= & i\partial_t\left( U^{-1} D_K(r,t;r,t) \right) -iD_K(r,t;r,t) \partial_t U^{-1}(t).
\end{align}

\section{Equivalence to Langevin dynamics} \label{noise}

Using the notation of Ref.~\onlinecite{Kamenevbook2011}, the bosonic quantum (\(\Delta_{q}\)) and classical
(\(\Delta_{c}\)) fields defined on the Keldysh contour, have the following correlators,
\begin{align}
  &D_R(1,2) = -i\langle \Delta_c(1)\Delta_q^*(2)\rangle,\nonumber\\
  &D_A(1,2) = -i\langle \Delta_q(1)\Delta_c^*(2)\rangle,\nonumber\\
  &D_K(1,2) = -i \langle \Delta_c(1) \Delta_c^*(2)\rangle,
\end{align}
above \(1,2\) denotes both spatial and temporal indices.
If, \(D_K = D_R\cdot \Pi_K \cdot D_A\), then the above correlators can equivalently be written
as a path integral over the bosonic fields as follows,
\begin{align}
  Z_K &= \int {\cal D}\left[\Delta_{c,q}, \Delta^*_{c,q}\right] e^{iS_K},\\
  S_K &= \int d1 \int d2 \begin{pmatrix} \Delta_c^* & \Delta_q^*\end{pmatrix}_{1} \begin{pmatrix}0 & D_A^{-1}\\D_R^{-1} & \Pi_K
    \end{pmatrix}_{1,2} \begin{pmatrix} \Delta_c \\ \Delta_q\end{pmatrix}_{2}.
\end{align}
According to Eq.~\eqref{Drdef}, and the line below,
\begin{align}
  D^{-1}_{R}(q,t,t') &= \frac{\nu}{T}\left(\partial_t + \lambda_q(t)\right)\delta(t-t');\nonumber \\
  \Pi_K(t,t') &= 4i\nu \delta(t-t').
\end{align}
One may introduce~\cite{Kamenevbook2011} an auxiliary field \(\xi\) that decouples the \(|\Delta_q|^2\) term in the action, (for
notational convenience we only highlight the time coordinate)
\begin{align}
  e^{-4 \nu\int dt \Delta_q \Delta_q^*} = \int{\cal D}\left[\xi,\xi^*\right] e^{-i\int dt \left[\xi \Delta_q^* + \xi^*\Delta_q\right]}
  e^{-\frac{1}{4\nu}\int dt \xi \xi^*}.
\end{align}
Writing the action in terms of this auxiliary field,
\begin{align}
  &S_K = \int d1 \int d2 \biggl[\Delta_q^* D_R^{-1}\Delta_c + \Delta_c^* D_A^{-1}\Delta_q\nonumber\\
&    -\xi \Delta_q^*
    - \xi^*\Delta_q +\frac{i}{4\nu}\xi\xi^* \biggr]. 
\end{align}
Requiring \(\delta S/\delta \Delta_q^*=0\) gives,
\begin{align}
D_R^{-1} \Delta_c = \xi,
\end{align}
where \(\langle \xi^*(t)\xi(t')\rangle = 4\nu \delta(t-t')\).
Giving the explicit expression for \(D_R\), and restoring the momentum label,
\begin{align}
\frac{\nu}{T}\left(\partial_t + \lambda_q(t)\right)\Delta_c(q,t) = \xi_q(t),
\end{align}
It is convenient to rescale \(\Delta_{c} \rightarrow \sqrt{\frac{\nu}{T}}\Delta_{c}\), then,
the Langevin equation is,
\begin{align}
\left(\partial_t + \lambda_q(t)\right)\Delta_c(q,t) = \sqrt{\frac{T}{\nu}}\xi_q(t),
\end{align}
implying that \(\Delta_c\) obeys Langevin dynamics where the noise is delta-correlated
with a strength equal to the temperature \(T\). 

\section{Derivation of kinetic coefficients}\label{Kder}
In this section we derive the kinetic coefficients assuming a rapid quench profile \(\lambda_q(t) = \theta(t) \lambda_q+ \theta(-t) r_i\),
where \(r_i/T = O(1)\). This condition simply ensures that the density of superconducting fluctuations is zero at \(t=0\).
The generalization
to general quench profiles is formally straightforward, and summarized in the main text.

\subsection{Derivation of Eq.~\eqref{KJ}}\label{KJder}
The term obtained from varying $n_k$ is,
\begin{align}
&K_J^{i}(t) \equiv \frac{1}{2}\text{Im}\int dt'\!\!\sum_{k,q}	iD_K(q,t,t')\nonumber\\
&\times	i\biggl[
		\delta n_{k+q/2}(t') + \delta n_{-k +q/2}(t')
	\biggr]
	\biggl[
		v^i_{k+q/2} + v^i_{-k + q/2}
	\biggr]\nonumber\\
&\times	e^{i\left(
		\epsilon_{k+q/2} + \epsilon_{-k+q/2}
	\right)(t-t')}.
\end{align}
Substituting Eq.~\eqref{DKdef} for $D_K$, and Eq.~\eqref{defn2} for \(\delta n_k\), we see that the above can be rewritten as
\begin{align}
&K_J^{i}(t) =  \int_0^t dt' \sum_q Q^{ij}_J(t,t') F_q(t')J^{j}(t'),
\nonumber\\
&Q^{ij}_J(t,t') \equiv
	\frac{\bar{m} T }{\rho \nu}\text{Im}\sum_{k}
	\biggl[
		\nabla^j_k n^{\rm eq}_{k+q/2}
		+
		\nabla^j_k n^{\rm eq}_{-k+q/2}
	\biggr]\nonumber\\
&\times 	
	i\left(
		v^i_{k+q/2}
		+
		v^i_{-k + q/2}
	\right)
	e^{i\left(
		\epsilon_{k+q/2} + \epsilon_{-k+q/2} + i\lambda_q
		\right)(t-t')
		}.		
\end{align}

Following the discussion in the main text, the term $Q^{ij}_{J}$ decays exponentially with rate $T$, whereas $J$ and $F$ change at a slower rate.
Therefore $Q_{J}$ appears like a delta function when integrated against $F(t)J(t)$, and we may write
\begin{align}
&K_J^i(t) \approx \sum_q F_q(t)J^j(t)\tilde{Q}^{ij}_J(t),\nonumber\\
&\tilde{Q}^{ij}_J(t)= \int_0^t dt'Q^{ij}_J(t,t') \approx
	-\frac{\bar{m} T }{\rho\nu}\text{Im} \sum_{k}\nonumber\\
&\times	\frac{
	\left(
		\nabla^j_k n^{\rm eq}_{k+q/2}
		+
		\nabla^j_k n^{\rm eq}_{-k+q/2}
	\right)
	\left(
		v^i_{k+q/2}
		+
		v^i_{-k + q/2}
	\right)
	 }{
		\epsilon_{k+q/2} + \epsilon_{-k+q/2} + i\lambda_q
		}
	\nonumber\\
	&\approx_{v_F q \ll T}\biggl(-
	\frac{\bar{m} T }{\rho\nu}\biggr)\text{Im} \sum_{k}
	\frac{
		q^r q^l m^{-1}_{il}\nabla^r_k\nabla^j_k n^{\rm eq}_{k}
	}{
		2\epsilon_{k} + i\lambda_q
		}
	\nonumber\\
	&\approx_{\lambda_q \ll T}
		\frac{\pi \bar{m} T}{\rho\nu} \sum_{k}
	\left(
		q^r q^l m^{-1}_{il}\nabla^r_k\nabla^j_k n^{\rm eq}_{k}
	\right)
	\delta(\epsilon_{k} + \epsilon_{-k})
\nonumber\\
	&= \frac{\pi \bar{m} T}{2\rho\nu} \sum_{k}
	\left(
		q^r q^l m^{-1}_{il}\frac{\partial^2\epsilon_k}{\partial_k^r\partial_k^j}\partial_{\epsilon_k}n^{\rm eq}_{k}
	\right)
	\delta(\epsilon_{k})
\nonumber\\
	&\approx
	\frac{\bar{m}\pi q^r q^l }{4  \rho}\langle m_{jr}^{-1} m_{il}^{-1} \rangle_{\rm FS}.
\end{align}
In the second line above we assume $t^{-1} \gg T$ so we are not too close to the quench. In the third line we take $q \ll  T/v_F$, and we also assume $\lambda_q/T \ll 1$.

\subsection{Derivation of Eq.~\eqref{KE}}\label{KEder}
Using Eq.~\eqref{DKdef} for $D_K$, and accounting for the change in $g_R$ due to the electric field in Eq.~\eqref{dEnk}, we find,
\begin{align}
&K_E^{i}(t) =  \int_0^t dt' \sum_q Q^{ij}_E(t,t')F_q(t') \frac{\rho}{\bar{m}}E^j(t'),
\nonumber\\
&Q^{ij}_E(t,t')E^j(t') \equiv-\text{Im}
	\frac{\bar{m}T}{\rho\nu} \sum_{k}
	\biggl(
		n^{\rm eq}_{k+q/2}
		+
	    n^{\rm eq}_{-k+q/2}
	\biggr)\nonumber\\
&\times	\biggl(
		v^i_{k+q/2}
		+
		v^i_{-k + q/2}
	\biggr)\nonumber\\
	&\times\int_{t'}^{t}\! ds\,v_k^j\left( A^j(t) - A^j(s)\right)
	e^{i\left(
		\epsilon_{k+q/2} + \epsilon_{-k+q/2} + i\lambda_q
		\right)(t-t')
		}.
\end{align}

Assume that the frequency of the electric field $\omega \ll T$ - in this case we may approximate the integral
\begin{equation}
\int_{t'}^{t} ds (A^j(t) - A^j(s)) \approx -E^j(t)(t-t')^2/2.
\end{equation}
Following the discussion in the main text, the term $Q^{ij}_{E}$ decays exponentially with rate $T$, whereas $J$ and $F$ change at a slower rate.
Therefore $Q_{E}$ appears like a delta function when integrated against $F(t)J(t)$, and we may write, as we did for $Q^{ij}_J$,
\begin{align}
K_E^i(t) &\approx \sum_q F_q(t)\frac{\rho E^j(t)}{\bar{m}}\tilde{Q}^{ij}_E(t),\nonumber\\
&\tilde{Q}^{ij}_E(t)=\int_0^t dt'Q^{ij}_E(t,t') \nonumber\\
&\approx
	\frac{\bar{m} T}{2\rho \nu}\text{Im}\sum_{k}\left(
		n^{\rm eq}_{k+q/2}
		+
		n^{\rm eq}_{-k+q/2}
	\right)\nonumber\\
&\times i	\frac{\left(
		v^i_{k+q/2}
		+
		v^i_{-k + q/2}
	\right)\left(
		v^j_{k+q/2}
		+
		v^j_{-k + q/2}
	\right)
	 }{(\epsilon_{k+q/2} + \epsilon_{-k+q/2} + i\lambda_q)^{3}}
	\nonumber\\	
	&\approx
	\frac{\bar{m} T}{\rho \nu}\text{Im}\sum_{k}i
	\frac{
		q^l q^r m^{-1}_{il} m^{-1}_{jr}n^{\rm eq}_{k}
	}{
		(2\epsilon_{k} + i\lambda_q)^3
		}
	\nonumber\\
 	&\approx
    \frac{\pi \bar{m} q^r q^l}{16 \rho}\langle m_{jr}^{-1} m_{il}^{-1} \rangle_{\rm FS} \int dx x^{-1} \frac{d^2}{d x^2 } \tanh(x/2)\nonumber\\
    &\approx
    \frac{28\pi \bar{m}  q^r q^l}{ 16\rho T}\zeta'(-2)\langle m_{jr}^{-1} m_{il}^{-1} \rangle_{\rm FS}.
\end{align}

\subsection{Derivation of Eq~\eqref{Jmem}} \label{Jmemder}

Substituting Eq.~\eqref{dellam1} into Eq.~\eqref{DR2}, and
assuming, $t > t'$, we obtain,
\begin{align}
i\delta D_K(t , t') &= -\nu  \int_0^{t'} ds D_R(t,s) D_A(s,t')\bigg[ \int^{t}_s du \delta \lambda_q(u) \nonumber\\
&+ \int^{t'}_s du \delta \lambda_q(u) \bigg] \\
	&= -2\nu\int^{t'}_0 du \delta \lambda_q(u) \int^u_0 ds D_R(t,s) D_A(s,t')\nonumber\\
&	 -\nu\int^t_{t'}du\delta \lambda_q(u) \int_0^{t'} ds D_R(t,s) D_A(s,t').
\end{align}
Using the definition of \(D_K= D_R\cdot \Pi_K \cdot D_A\) with \(\Pi_K(t,t')= 4\nu\delta(t-t')\), the above becomes,
\begin{align}
i\delta D_K(t , t') &=-\frac{1}{2}\int^{t'}_0 du e^{-\lambda_q(t'-u)}\delta \lambda_q(u) i D_K(t,u)\nonumber\\
&	 -\frac{1}{4}\int^t_{t'}du\delta \lambda_q(u) iD_K(t,t').
\end{align}
Then we use Eq.~\eqref{DKdef} to write,
\begin{widetext}
\begin{align}
&i\delta D_K (t> t') = -\frac{2T}{2\nu}e^{-\lambda_q(t-t')}
\int_0^{t'}\!du\, e^{-2\lambda_q(t' - u)}  \delta \lambda_q(u) F_q(u)
-\frac{T}{2\nu} e^{-\lambda_q(t-t')} \int_{t'}^t du\delta\lambda_q(u) F_q(t')\nonumber\\
&=
	-\frac{2T}{2\nu} e^{-\lambda_q(t-t')}
		\int_0^{t}\!du\, e^{-2\lambda_q(t' - u)}
		\delta \lambda_q(u) F_q(u)
-\frac{T}{2\nu} e^{-\lambda_q(t-t')}
    	\int_{t'}^t du
    	\delta\lambda_q(u) \left(F_q(t') - 2e^{-2\lambda_q(t' - u)}F_q(u)\right)
\nonumber\\
&=
	-\frac{2T}{2\nu} e^{\lambda_q(t-t')}
		\int_0^{t}\!du\, e^{-2\lambda_q(t - u)}
		\delta \lambda_q(u) F_q(u)
   -\frac{T}{2\nu} e^{-\lambda_q(t-t')}
    	\int_{t'}^t du
    	\delta\lambda_q(u) \left(F_q(t') - 2e^{-2\lambda_q(t' - u)}F_q(u)\right)
\nonumber\\
&\approx -\frac{2T}{2\nu} e^{\lambda_q(t-t')}
		\int_0^{t}\!du\, e^{-2\lambda_q(t - u)}
		\delta \lambda_q(u) F_q(u)
    +\frac{T}{2\nu}e^{-\lambda_q(t-t')} (t-t')
    	\delta\lambda_q(t)F_q(t),
\end{align}
\end{widetext}
where in the last line we used the fact that $t-t'$ is of order $T^{-1}$ as \(iD_K\) appears
along with \(\Pi_A\). We discuss this last term further below and
show it to be parametrically smaller than the local terms calculated earlier.

We insert the second part into the collision integral giving,
\begin{widetext}
\begin{align}
&4\int_0^t dt'\text{Im} \Pi'_A(t',t)
	\left(
		e^{-\lambda_q(t-t')}
		\frac{T}{\nu} (t-t') \delta \lambda_q(t) F_q(t)
	\right)
		 \propto
	\delta\lambda_q(t) F_q(t)
	\text{Im}\sum_k
	i \frac{
		n^{\rm eq}_{k+q/2} + n^{\rm eq}_{-k+q/2}
	}{\
		(\epsilon_{k+q/2} +\epsilon_{-k +q/2} - i\lambda_q)^2
	}
\nonumber\\
	&\approx_{q\rightarrow 0}
	\delta\lambda_q(t) F_q(t)
	\text{Im}	
	\sum_k i
	\frac{
		2n^{\rm eq}(\epsilon_k)
	}{
		(2\epsilon_k - i \lambda_q)^2
	}
	\approx
	\delta\lambda_q(t) F_q(t)
	\text{Im}	
	i\int d\epsilon \, \nu
	\frac{
		2n^{\rm eq}(\epsilon)
	}{
		(2\epsilon - i \lambda_q)^2
	}
\nonumber\\
	&\approx_{\lambda_q\rightarrow 0}
	\delta\lambda_q(t) F_q(t)
	\text{Re}	
	\int d\epsilon \, \nu
	\mathcal{P}\frac{
		\partial_\epsilon n^{\rm eq}(\epsilon)
	}{
		\epsilon
	} = 0.
\end{align}
\end{widetext}
Thus this term is parametrically smaller than the local term already calculated.

Now we consider the full term
\begin{widetext}
\begin{align}
&4\int_0^{t} dt' \text{Im}\biggl[v_k^i i \delta D_K(t,t') \Pi'_A(t',t)
\biggr]\nonumber\\
&=  \frac{1}{2}\text{Im}\int_0^t dt'
 \sum_{k,q}i
 \left(v^i_{k+q/2} + v^i_{-k+q/2}\right) i\delta D_K(t',t)\left(n^{\rm eq}_{k+q/2}+ n^{\rm eq}_{-k+q/2}\right)e^{i(\epsilon_{k+q/2} + \epsilon_{-k+q/2})(t-t')}\biggr]\nonumber\\
& \approx
 \text{Im} \biggl[\frac{T}{2\nu}
 \sum_{k,q} i\int_0^t dt'
 \left(v^i_{k+q/2} + v^i_{-k + q/2}\right)\left(n^{\rm eq}_{k+q/2}
+ n^{\rm eq}_{-k+q/2}\right)
e^{i(\epsilon_{k+q/2} + \epsilon_{-k+q/2} -i \lambda_q)(t-t')}\nonumber\\ &\times \int_0^{t} du\,
\left(-
 e^{-2(t-u)\lambda_q}
 \delta\lambda_q(u) F_q(u)
 \right)\biggr]
\nonumber \\
&\approx
 \text{Im}\biggl[\frac{T}{2\nu}\sum_{k,q}
 q^j m^{-1}_{ij}
 \frac{n^{\rm eq}_{k+q/2} + n^{\rm eq}_{-k+q/2}}
      {\epsilon_{k+q/2} + \epsilon_{-k+q/2} - i\lambda_q}
\int_0^t du
 \,e^{-2(t-u)\lambda_q}
 \delta \lambda_q(u) F_q(u)\biggr]
\nonumber\\
& \approx\sum_q
	\frac{ q^l T}{2\nu}
	\sum_k m^{-1}_{il}
	\frac{n^{\rm eq}(\epsilon_k) \lambda_q}
	     {4\epsilon_k^2 + \lambda_q^2}
\int_0^t du
    e^{-2(t-u)\lambda_q}
    \delta \lambda_q(u) F_q(u)
\nonumber\\
& = \sum_q
	\frac{T q^l }{2\nu}
	\left(
		\sum_k m^{-1}_{il}
		\frac{n^{\rm eq}(\epsilon_k) \lambda_q}
		     {4\epsilon^2_k + \lambda_q^2}
	\right)\int_0^t du
	e^{-2(t-u)\lambda_q}
	F_q(u)
	\frac{2 q^j\bar{m}}{\rho M}
	\left[
		J^j(u) +
		\int_u^t dt'
		\frac{\rho E^j(t')}
			 {\bar{m}}
	\right]\nonumber\\
&= \sum_q\alpha q^i q^j \lambda_q
	\int_0^t du
	e^{-2(t-u)\lambda_q} F_q(u)\left[
		J^j(u)+
		\int_u^t dt'
		\frac{\rho E^j(t')}
			 {\bar{m}}
	\right].
\end{align}
\end{widetext}
Above, in the second last line, we used the expression for \(\delta \lambda_q\) in Eq.~\eqref{dellam1}.

\section{Derivation of Eq.~\eqref{hder}}\label{hderapp}

Let us now discuss the limiting behaviors of \(h(z_0)\).
If $z_0 \ll 1 $, the integral is restricted to the region $\bar{x} \ll 1$ and we may approximate
$\left(1 - e^{-\bar{x}}\right)/\bar{x} \approx 1$. Therefore in this limit we have $\text{Re}\left[h\right]\approx 2z_+^0$.
If $|z_+^0| \gg 1$ then we split the integral at  $\bar{x} = 1$. The lower part
\begin{align}
\int_0^1 \! d\bar{x}\,\frac{1- e^{-\bar{x}}}{\bar{x}}\left[\left(z_+^0\right)^2 - \bar{x} \right]^{-1/2}
&\sim
\int_0^1 \! d\bar{x}\,\frac{1- e^{-\bar{x}}}{\bar{x}}\frac{1}{z_+^0}\nonumber\\
&\times \left( 1 + \frac{\bar{x}}{2\left( z_+^0\right)^2} +\cdots\right) \nonumber\\
&\propto \frac{1}{z_+^0 } + \mathcal{O}\left(\left( z_+^0\right)^{-2}\right).
\end{align}
Whereas in the part from 1 to $\infty$ the exponential may be neglected leaving the leading part
\begin{align}
\int_1^\infty \! d\bar{x}\,\frac{1}{\bar{x}}\left[\left(z_+^0\right)^2 - \bar{x} \right]^{-1/2}
&\sim \frac{\log \,\left(z_+^0\right)^2}{z_+^0}.
\end{align}
Thus we have Eq.~\eqref{hder}.

\section{Derivation of Eq.~\eqref{selfnu}}\label{nuder}
We record here some facts about $f(z)$ as a  function in the complex plane. It may be written as
\begin{equation}
f(z) = \frac{1}{\sqrt{z^2 +1}}\left[ \log\left( \frac{\sqrt{z^2 + 1}  + z }{\sqrt{z^2 + 1}  - z }\right) - \pi i \text{sgn}\Im z\right].
\end{equation}
It is analytic in the upper and lower half planes separately with a branch cut along the real axis.
The points $z = \pm i$ which appear singular are in fact smooth. One may also convince one self that
$|f(z)|\leq \pi$ for all $z$, reaching this limit only when $z = 0\pm i \delta$.
It is also an injective function of $z$. In addition, we have that as
$z\rightarrow \infty$,
\begin{equation}
f(z) \sim 2\frac{\log z}{z},
\end{equation}
and that $f$ is pure imaginary on the imaginary axis.
Causality requires that $z_\pm$ have no zeros or branch cuts in the upper half complex $\omega$ plane,
but there will be analytic structure as a function of $\varepsilon$.

Let us first test the validity of the perturbative approximation. To zeroth order we set $z_\pm = z_\pm^{0}$.
To first order we substitute this into the rhs of equation~\eqref{SRa},~\eqref{SAa}, and obtain,
\begin{equation}
z_\pm = z_\pm^0 + \gamma f\left(z_\mp^0\right),
\end{equation}
and to second order
\begin{align}
z_\pm &= z_\pm^0 + \gamma f\left(z_\mp^0 + \gamma f\left(z_\pm^0\right)\right)\nonumber\\
&\approx z_\pm^0 + \gamma f\left(z_\mp^0\right) + \gamma^2 f'\left(z_\mp^0\right)f\left(z_\pm^0\right)\label{pert1}.
\end{align}
Comparing the first and second correction we see that the latter is smaller when
\begin{equation}
\gamma \ll \left|\frac{f\left(z_\mp^0\right)}{f'\left(z_\mp^0\right) } \frac{1}{f\left(z_\pm^0\right)}\right|.
\end{equation}
Considering the various limits, we see that the rhs is always at least $O(1)$ so the perturbative treatment is always valid if $\gamma \ll 1$.
Making the very coarse approximation $|f| \sim \min(1,|z^{-1}|)$, we get roughly
\begin{equation}
\gamma \ll \min(1,|z_+^0|)\cdot \min(1,|z_-^0|).
\end{equation}
This may still be satisfied even if $\gamma \geq 1$. If $\omega/T \gg 1/\text{Gi}$, the perturbative limit will hold for all $\varepsilon$. With the weaker condition $\omega^2/T^2 \gg \text{Gi}$, the perturbative limit holds except in the narrow region $|\varepsilon \pm \omega|/T \ll \sqrt{\epsilon} $, which gives only a small contribution to $\delta \nu(\omega)$.

Now let us consider the behavior in the simplest case when $z_\pm^0 = 0$. In this case we have a self-consistent solution when $z_\pm = \mp i y$, $y$ real, $y>0$, since this is equivalent to finding a solution to
\begin{equation}
-i y(\gamma) =   \gamma  f(  i y(\gamma)),
\end{equation}
which is possible since $f$ is imaginary on the imaginary axis. In the limiting case of $\gamma \gg 1 $, we get $y^2 \sim \gamma \log y^2$ which gives to logarithmic accuracy,
\begin{align}
y \sim \sqrt{\gamma \log \gamma}\label{ydef}.
\end{align}
In the limit $\gamma \ll 1$, using the expansion of $f(z)$ for small argument, we obtain $z_{\pm } \approx \mp i\gamma \pi + \mathcal{O}(\gamma^2)$.

Now let us advance to the case $z_\pm^0 = c \in \mathbb{R}$. Recall $\omega=0$. This has a self-consistent solution when $z_+ = z = z_-^*$,
\begin{equation}
z - c= \gamma f(z^*),
\end{equation}
since $f(z)^* = f(z^*)$. The pertubative condition in this case is
\begin{equation}
\gamma \ll 1/|f'(c)| \sim c^2/ \log|c^2|.
\end{equation}
Thus the perturbative regime begins when $c \gg y(\gamma)$. In this limit,
we obtain the usual perturbative result by setting $z\approx c$.

In the opposite limit, let us assume $|z|\gg 1,|c|$. On these assumptions,
writing $z = -i|z|e^{-i\phi}$, and comparing real and imaginary parts of
\begin{align}
\left(z- c\right) z^* &=  \gamma \log(-(z^*)^2) \nonumber\\
\Rightarrow |z|^2  - i c |z|e^{i\phi} &= 2\gamma\left[\log|z| + i \phi\right],\nonumber
\end{align}
we obtain,
\begin{align}
\qquad |z|^2 + c |z| \sin \phi &=  \gamma \log |z|^2 ,\\
\text{\&}\qquad - c |z|\cos \phi &= 2\gamma \phi.
\end{align}
On the assumption $|z| \gg |c|$ the first line above gives $|z| = y(\gamma)$. The second line can be rearranged to give,
\begin{equation}
\frac{\phi}{\cos{\phi}}  = -c \frac{y(\gamma)}{2\gamma}.
\end{equation}
Thus $\phi$ goes from $\pi/2$ to $-\pi/2$ as $c$ goes from $-\infty$ to $\infty$. We may split this into two limits. When $c \ll \gamma / y(\gamma) = y(\gamma)/\log y(\gamma)$,
we have that $\phi \ll 1$ and so the $\cos$ may be set to $1$ giving $\phi = -c y(\gamma)/ 2 \gamma$, and thus
\begin{align}
z &= -i|z| (1 - i \phi)
= -i y(\gamma)  +  c \log(\gamma)/2.
\end{align}
In the other limit, $cy(\gamma)/\gamma \gg 1$ is large so $\phi$ is close to $\pm \pi/2$. Let us take $c> 0$, then
\begin{gather}
\phi = -\frac{\pi}{2}  + \frac{\pi \gamma}{c y(\gamma)}.
\end{gather}
so that $z$ is given by,
\begin{gather}
 z = -i e^{i \pi/2} |z|\left( 1 - i\frac{\pi \gamma}{c y(\gamma)} \right)\nonumber\\
= y(\gamma) - i \frac{\pi\gamma}{c}.
\end{gather}
This is almost identical to the perturbative calculation with a slight correction to the real part.

We can collect all of these limits,
\begin{equation}
z= \begin{cases}
   -i y(\gamma)  +  c \log(\gamma)/2 & c \ll y(\gamma)/\log(\gamma) \\
   y(\gamma) - i \frac{\pi\gamma}{c} & y(\gamma)/\log(\gamma) \ll c \ll y(\gamma)\\
  c +  \frac{\gamma}{c}\left[\log{c^2} - \pi i\right]  & y(\gamma) \ll c
\end{cases}
\end{equation}
where in the last case, we have used the perturbative expansion in Eq.~\eqref{pert1}.

We are now in a position to estimate the density of states for $\omega = 0$, $\gamma \gg 1$, and $c = \varepsilon/T\sqrt{r}$. Starting from
\begin{equation}
\frac{\nu (\omega)}{\nu_0} = \frac{1}{\pi}\Im \int_{-\infty}^{\infty}
d\varepsilon  \frac{1}{\sqrt{r}Tz\left(\omega = 0,\varepsilon\right)}.
\end{equation}
Combining the positive and negative integrals over $\varepsilon$ or $c$,
\begin{equation}
\frac{\nu (\omega)}{\nu_0} = \frac{1}{\pi}\Im \int_{0}^{\infty}
d c\left[\frac{1}{z(c)} + \frac{1}{z(-c)}\right].
\end{equation}
Next use the fact that at $\omega=0$, $z(c) = -z(-c)^*$, which follows from the above calculation. This can also be seen
from noting that $f(-z)=-f(z)$. Then, using $z_+(c)=c+i\delta +\gamma f(z_-(c))=c+i\delta +\gamma f(z_+^*(c))$. Taking $c\rightarrow -c$ and
conjugating, this becomes, $z_+^*(-c)= -c-i\delta +\gamma f(z_+(-c))=-c-i\delta -\gamma f(-z_+(-c))$.
Thus $z_+(-c)^*=-z_+(c)$ is a solution of the above equation as this simultaneously requires $f(-z_+(-c))= f(z_+^*(c))$.
Using this gives,
\begin{equation}
\frac{\nu (\omega=0)}{\nu_0} = \frac{2}{\pi} \int_{0}^{\infty}
d c\left[ \frac{\Im z(c)}{[\Re z(c)]^2 + [\Im z (c)]^2 }\right].
\end{equation}
Now we split this integral into three regions, according to the asymptotics above,
\begin{equation}
\int_{0}^{\infty}
d c\left[ \frac{\Im z(c)}{[\Re z(c)]^2 + [\Im z (c)]^2 }\right] = I_1 + I_2 + I_3,
\end{equation}
where,
\begin{align}
I_1 \equiv & \int_{0}^{y(\gamma)/\log\gamma}
d c\left[ \frac{\Im z(c)}{[\Re z(c)]^2 + [\Im z (c)]^2 }\right] \\
 & = \int_{0}^{y(\gamma)/\log\gamma}
d c\left[ \frac{y(\gamma)}{[c\log(\gamma)/2]^2 + [y(\gamma)]^2 }\right], \nonumber\\
&\qquad \text{define } u \equiv c \frac{\log\gamma }{y(\gamma)}, \nonumber\\
&I_1 = \frac{1}{\log{\gamma}}\int_{0}^{1}
 d u\left[ \frac{1}{(u/2)^2 + 1 }\right] \propto \frac{1}{\log{\gamma}}.
\end{align}

\begin{align}
I_2 \equiv  & \int_{y(\gamma)/\log y(\gamma)}^{y(\gamma)} d c\left[ \frac{\Im z(c)}{[\Re z(c)]^2 + [\Im z (c)]^2 }\right] \\
& = \int_{y(\gamma)/\log \gamma}^{y(\gamma)} d c\left[ \frac{\pi \gamma /c }{y(\gamma)^2 + [\pi\gamma/c]^2 }\right], \nonumber\\
&\qquad  \text{define  } u = c \cdot y(\gamma)/\gamma \approx c\dot\log(\gamma)/y(\gamma), \nonumber \\
& I_2= \frac{\gamma}{y(\gamma)^2}\int_{1}^{\log y(\gamma)} d u\left[ \frac{\pi /u }{1 + [\pi/u]^2 }\right]\nonumber\\
& \approx \frac{\gamma}{y(\gamma)^2} \int_1^{\log y(\gamma)} du \frac{\pi}{u}\propto \frac{1}{\log \gamma } \log \log y(\gamma)\nonumber\\
& \propto \frac{1}{\log \gamma}.
\end{align}
Here we are treating $\log\log \gamma $ as an $\mathcal{O}(1)$ constant, which has been the order of our precision throughout.
\begin{align}
I_3 \equiv & \int_{y(\gamma) }^{\infty}
d c\left[ \frac{\Im z(c)}{[\Re z(c)]^2 + [\Im z (c)]^2 }\right] \\
& =\int_{y(\gamma) }^{\infty}
d c\left[ \frac{\gamma \pi / c}{c^2 + [\gamma \pi/ c]^2 }\right] \nonumber\\
& \approx \int_{y(\gamma)}^{\infty}
	d c \frac{\gamma \pi }{c^3}= 2\pi \frac{\gamma}{y(\gamma)^2}=  \frac{2\pi}{\log{y(\gamma)}} \nonumber\\
&\approx \frac{4\pi}{\log{\gamma}}.
\end{align}

Thus we see that all terms contribute $\propto 1/ \log{\gamma}$ and therefore we have that
\begin{equation}
\nu(\omega = 0) \propto \frac{\nu_0}{\log \left(\text{Gi}/8\pi r\right)}.
\end{equation}

\end{appendix}


%

\end{document}